\documentclass[aps,prb,twocolumn,showpacs,nobibnotes]{revtex4}
\usepackage{graphics,graphicx,amsfonts,amsmath,amsbsy,amssymb,color}
\usepackage{bm}
\usepackage{subfig}
\usepackage{psfrag}  
\usepackage{multirow}

\def \bfr {{\bf r}}

\def \bfi {{\bf i}}
\def \bfj {{\bf j}}
\def \bfk {{\bf k}}

\def \bfk {{{\bf k}}}

\def \bfg {{\bf g}}

\def \bfx {{\bf x}}

\def \beq {\begin{eqnarray}}
\def \eeq {\end{eqnarray}}

\def \Schrodinger {{Schr\"{o}dinger }}

\def \iFCIQMC {{\mbox{\emph{i}-FCIQMC }}}
\def \iFCIQMCbracket {{\mbox{\emph{i}-FCIQMC}}}

\def \bfj {{\bf j}}
\def \bfi {{\bf i}}

\newcommand{\appropto}{\mathrel{\vcenter{
  \offinterlineskip\halign{\hfil$##$\cr
    \propto\cr\noalign{\kern2pt}\sim\cr\noalign{\kern-2pt}}}}}

\def \ket {{\rangle}}
\def \bra {{\langle}}
\newcommand{\Hamil}{\hat{H}}

\newcommand{\vft}[1]{v_{#1}}

\newcommand{\refeq}[1]{{Eq.~\ref{#1}}}
\newcommand{\reffig}[1]{{Fig.~\ref{#1}}}
\newcommand{\refsec}[1]{{Sec.~\ref{#1}}}
\newcommand{\reftab}[1]{{Table \ref{#1}}}
\newcommand{\half}{\frac{1}{2}}

\begin{document}
\title{Investigation of the Full Configuration Interaction Quantum Monte Carlo Method Using Homogeneous Electron Gas Models}
\author{James~J.~Shepherd}
\email{js615@cam.ac.uk}
\affiliation{University of Cambridge, Chemistry Department, Lensfield Road, Cambridge CB2 1EW, U. K.}
\author{George~H.~Booth}
\affiliation{University of Cambridge, Chemistry Department, Lensfield Road, Cambridge CB2 1EW, U. K.}
\author{Ali~Alavi}
\email{asa10@cam.ac.uk}
\affiliation{University of Cambridge, Chemistry Department, Lensfield Road, Cambridge CB2 1EW, U. K.}
\begin{abstract}
Using the homogeneous electron gas (HEG) as a model, we investigate the sources of error in the `initiator' adaptation to Full Configuration Interaction Quantum Monte Carlo (\iFCIQMCbracket), with a view to accelerating convergence. In particular we find that the fixed shift phase, where the walker number is allowed to grow slowly, can be used to effectively assess stochastic and initiator error. Using this approach we provide simple explanations for the internal parameters of an \iFCIQMC simulation. We exploit the consistent basis sets and adjustable correlation strength of the HEG to analyze properties of the algorithm, and present finite basis benchmark energies for $N=14$ over a range of densities $0.5 \leq r_s \leq 5.0$~a.u. A \emph{single-point extrapolation} scheme is introduced to produce complete basis energies for 14, 38 and 54 electrons. It is empirically found that, in the weakly correlated regime, the computational cost scales linearly with the plane wave basis set size, which is justifiable on physical grounds. We expect the fixed shift strategy to reduce the computational cost of many \iFCIQMC calculations of weakly correlated systems. In addition, we provide benchmarks for the electron gas, to be used by other quantum chemical methods in exploring periodic solid state systems.
\end{abstract}
\date{\today}
\maketitle


\section{Introduction}

The simulation-cell homogeneous electron gas (HEG), consisting of $N$ electrons in a finite and periodic box of length $L$, with a uniform neutralizing background, should be a compelling choice of model system for quantum chemical studies. Any determinant comprised of plane waves is an exact Hartree-Fock solution as well as the exact natural orbital representation for the gas\cite{Davidson}, and Hamiltonian matrix elements are analytically computable\cite{Vignale}. A single tunable density parameter ($r_s$) controls the strength of coupling, and the gas is then representative of wide range of weakly to strongly correlated electronic problems\cite{Vignale}.

The complete one-particle space comprises of an infinite set of plane waves, and a common choice of wavevectors in a finite basis is a gridded sphere centred on the origin of reciprocal space. Since it provides a basis set tunable with only one parameter, a reciprocal space radial cutoff $k_c$, the limit $k_c\rightarrow\infty$, corresponding to the complete basis set limit, can be approached systematically and straightforwardly, with no need to re-optimise the orbitals between changes in basis set, since the Fock operator does not couple any spin orbitals. 

In spite of these apparent advantages, there has been little investigation of HEG using quantum chemical methods. Perhaps one drawback preventing such work is that expectation values of such a finite $N$-electron gas differ from that of the infinite system and as such the thermodynamic limit, $N\rightarrow\infty$ with the density held constant, needs to be found to converge on physical characteristics accurately. Small-$N$ simulation-cell gases suffer from so-called finite size effects, which can produce non-physical behavior in simulations\cite{Drummond2008,Lin}. Nevertheless, finite size corrections have been developed that allow physical electron-gas behavior to be observed for increasingly low electron numbers\cite{Drummond2008,Holzmann2011,Fraser1996,Drummond2009,Krakauer2008}.

The `exact' solution to the electronic \Schrodinger equation for a finite basis can be solved by expanding the wavefunction as an optimized linear combination of all Slater Determinants that can be formed from rearranging $N$ electrons in $M$ basis functions. If these are found by an exact diagonalization, this method is referred to as full configuration interaction (FCI) and scales combinatorially in $N$ and $M$ (Ref. \onlinecite{Knowles1984}). Truncated CI techniques, restricting the calculation to a subset of the space, although potentially polynomially scaling, would yield zero correlation energy per electron in the thermodynamic limit due to lack of size extensivity\cite{Langhoff}. This makes treatment of the HEG of a modest size in even a tiny basis set intractable.

FCI Quantum Monte Carlo and its `initiator' adaptation (\iFCIQMCbracket) are novel methods developed in a series of recent papers\cite{FCIQMCPaper1,FCIQMCPaper2,c2paper,IPPaper,EApaper,UEGPaper1} in which the FCI equations are simulated by representing the determinant coefficients as a set of walkers evolving over discretized imaginary time. This allows much larger Hilbert spaces to be studied, with the largest space accurately sampled to date being $10^{108}$, in a previous study of the 54-electron gas\cite{UEGPaper1}. In this study, the high-density gas was explored using \iFCIQMC to yield energies of, in principle, FCI accuracy. The error incurred by using a finite basis was removed by an extrapolation scheme proposed by the authors and to be expanded on in a forthcoming paper. Comparison between our energies and those of recent diffusion Monte Carlo (DMC) calculations, based on a similar methodology as the famous study of Ceperley and Alder\cite{CeperleyAlder1980}, were consistent with the claim that modern DMC energies for finite electron gases are thought to be accurate to within 1m$\text{E}_\text{h}$ per electron\cite{Rios,Kwon1998}. 

It is our intention to continue to explore the use of \iFCIQMC to study the electron gas. In this paper we seek to use the advantages presented by the HEG to better-understand the method itself, and its potential application to periodic systems. In particular, we show that the only approximation made in \iFCIQMC, arising from using a finite number of walkers, is rigorously controllable and can be removed in a systematic fashion with the use of the fixed shift strategy. This is a minor adaptation to the current algorithm to achieve FCI accuracy reliably, and comparatively cheaply compared to a previous study. In doing so, we will also expose some of the benefits of using the HEG for studies using quantum chemical methods, and provide much-needed literature benchmarks.

\section{FCIQMC}
\label{sec_FCIQMC}

We seek to find the ground state wavefunction and energy of the $N$-electron HEG in a simulation cell with periodic boundary conditions in the plane-wave representation. The single particle states are given by,
\begin{equation}
\psi_{j}  (\bfx) \equiv \psi_{j}  (\bfr , \sigma) =\sqrt{\frac{1}{\Omega}}~e^{i \bfk_j \cdot \bfr} ~\delta_{\sigma_j,\sigma},
\end{equation}
specified by a set of reciprocal lattice vectors $\{\bfk_j\}$, where $\Omega$ is the real-space unit cell volume of a cubic cell. Imposing a cubic symmetry to the simulation cell allows $\bfk$ to take values of $\frac{2\pi}{L} \left(n,m,l\right)$ where $n$,$m$ and $l$ are integers. We then use a single cutoff parameter, $k_c$, to confine our basis set to be the $M$ spin orbitals resulting from those plane waves of a kinetic energy less than $\half k_c^2$.

The simulation-cell HEG Hamiltonian can be written using second quantization as:
\begin{equation}
\begin{split}
\Hamil=&\sum_{ij}  t_{i}^{j} a_i^\dagger a_j  + \half \sum_{ijkl} v_{ij}^{kl} a_i^\dagger a_j^\dagger a_l a_k + \half \sum_i v_\text{M} a_i^\dagger a_i,
\end{split}
\end{equation}
where $i$, $j$, $k$ and $l$ refer to single-particle plane waves. The $t_{i}^{j}$ matrix elements are due to the kinetic energy operator,
\begin{equation}
 \begin{split}
 t_{i}^{j}&= -\half \bra i | \nabla_1^2 | j \ket \\
 &= \half \bfk_i^2 \delta_{ij},
 \end{split}
\end{equation}
which is diagonal in the plane wave representation. The two-particle operator, containing electron-electron interactions, electron-background interactions and the background-background interaction, is represented by,
\begin{equation}
 \begin{split}
v_{ij}^{ab}= \vft{\bfg}  \delta_{\bfg,\bfk_a-\bfk_i} \delta_{\bfg,\bfk_j-\bfk_b},
 \end{split}
\end{equation}
where, 
\begin{equation}
\vft{\bfg} = \left\{
\begin{array}{ll}
\frac{1}{\Omega} \frac{4\pi}{\bfg^2}, & \bfg\neq\bf{0} \\
0, & \mbox{\bfg=\bf{0}}
\end{array}
\right. 
\label{VfourT}
\end{equation}
and where $\bfg$ is the change in the one-particle momentum due to the excitation $ij\rightarrow ab$. The remaining term, $v_\text{M}$, is the Madelung term, which represents contributions to the one-particle energy from interactions between a point charge and its own images and a neutralising background\cite{Ewald,Fraser1996}. This is an artifact of performing a simulation-cell calculation and vanishes in the thermodynamic limit. The term in $v_\text{M}$ also cancels between the total FCI energy and the Hartree-Fock energy, making the (FCI) correlation energy independant of its value.

The FCI solution to the \Schrodinger equation expressed in a basis of spin orbitals can be written as an optimized linear combination of Slater Determinants,
\begin{equation}
\Psi = \sum_\bfi C_\bfi | D_\bfi \ket,
\label{SDexp}
\end{equation}
which are antisymmetrized products of $N$ normalized spin-orbitals,
\begin{equation}
D_\bfi = \mathcal{A} \left[ \psi_i(\bfx_1) \psi_j(\bfx_2) ... \psi_k(\bfx_N) \right].
\end{equation}
All determinants formed from the rearrangement of the $N$ electrons in the $2M$ spin orbitals are included in the sum over $\bfi$, which uniquely labels each determinant\cite{Kutzelnigg1997}. In the FCI approach to this problem, the coefficients are found by diagonalization of the Hamiltonian matrix.

In a recently developed quantum Monte Carlo algorithm, termed Full Configuration Interaction QMC (FCIQMC)\cite{FCIQMCPaper1}, the ground state wavefunction and energy are found by a long-time integration,
\begin{equation}
\Psi_0=\lim_{\tau\rightarrow\infty}e^{-\tau\left(\Hamil-E_0\right)} D_{\bf{0}}.
\end{equation}
This can be re-cast in terms of the Hamiltonian matrix as a set of coupled equations for the determinant coefficients
\begin{equation}
-\frac{dC_\bfi}{d\tau} = (H_{\bfi\bfi}-E_{\text{HF}}-S)C_{\bfi} + \sum_{\bfj \neq \bfi} H_{\bfi\bfj}C_{\bfj},
\label{FCI-eqs}
\end{equation} 
where $E_{\text{HF}}$ is the Hartree-Fock energy and an arbitrary energy `shift', $S$, has been introduced. These equations are then regarded as a set of master equations governing the dynamics of the evolution of the determinant coefficients in imaginary time, with elements of $\bf{H}$ being non-unitary transition rates. The sign problem in this form of quantum Monte Carlo is generally ameliorated compared with that of diffusion Monte Carlo\cite{Spencer}.

These dynamics are simulated by introducing a population of $N_w$ `walkers', which, when distributed over the determinants, represent the sign of the coefficients in the FCI expansion for the purposes of the simulation,
\begin{equation}
C_\bfi \propto \bra N_\bfi \left(\tau\right) \ket,
\end{equation}
where each walker can have a positive or negative sign. The walker population is then allowed to evolve through discretized imaginary time-steps by spawning, death/cloning and annihilation events according to \refeq{FCI-eqs} until a steady-state is reached. 

The exact rules for this can be found in Ref. \onlinecite{FCIQMCPaper1}, but are described briefly here:
\begin{enumerate}
\item In the spawning step, each walker is considered in turn. A connected determinant $D_\bfj$ is chosen with a normalized probability $p_{\text{gen}}(\bfj|\bfi)$, and an attempt is made to spawn onto this determinant with probability $\frac{\delta\tau |H_{\bfi\bfj}| } {p_{\text{gen}}(\bfj|\bfi)}$. Attempts are generally restricted to coupled determinants, defined by $H_{\bfi\bfj}$ being non-zero, for efficiency. If this value exceeds 1, the number of walkers spawned is related to the amount by which this value exceeds 1. The sign is determined as the same as the parent if $H_{\bfi\bfj}<0$ and the opposite sign otherwise. 
\item In the death/cloning step each walker attempts to die or clone itself with probability $\delta\tau (H_{\bfi\bfi} -E_\text{HF}-S)$ where the walker dies if this is positive and is cloned if this is negative. 
\item Finally, in the annihilation step each walker is considered and removed if there is an opposite-signed walker at the same determinant.
\end{enumerate}

The simulation has two phases:
\begin{enumerate}
\item \emph{Fixed shift mode.} In this period of the calculation, the shift ($S$) is fixed at a constant value. This should result in an exponential growth of walkers as long as $S$ is greater than the correlation energy, whose rate depends on the value of this shift, the timestep and the correlation energy of the problem. Increasing the value that the shift is fixed at relative to the correlation energy will result in faster growth, however it has been observed in some cases that this can result in longer equilibration times once the shift is allowed to vary.
\item \emph{Variable shift mode.} When a target walker number has been reached the simulation proceeds to vary $S$ to keep the walker number $N_w$ constant. After an equilibration period, for high enough $N_w$, the determinant populations equilibrate to a distribution proportionate to the FCI wavefunction. The parameter $S$ therefore is a population control parameter. 
\end{enumerate}

The energy can be found in two ways from the simulation. In variable shift mode, $S$ is updated self-consistently at equilibrium and oscillates around the correlation energy as expected from \refeq{FCI-eqs}. 

However, throughout this work, the projected energy is used as an energy estimator for the dynamic,
\begin{equation}
E_{\text{FCIQMC}} = \lim_{\tau\rightarrow\infty} \sum_{\bfj} \bra D_\bfj | H | D_{\bf 0} \ket \frac{\bra N_{\bfj} (\tau)\ket}{\bra N_{\bf 0} (\tau)\ket},
\label{projEeq}
\end{equation}
where $D_{\bf 0}$ is taken as the Hartree-Fock determinant and $\bfj$ is taken as a sum over doubly-excited determinants.

Typically the walker population is initially grown by setting $S$ equal to zero, from one walker on the HF determinant. Only populations above a critical system-dependent size are able to converge to the FCI distribution, and this size was found to scale linearly with the size of the Hilbert space\cite{FCIQMCPaper1}. Nevertheless, small prefactors to this scaling allowed the method to be used to achieve FCI accuracy on a range of systems which were previously out of reach of traditional diagonalization algorithms\cite{IPPaper}.

However, in order to alleviate this scaling problem, an adaptation of this method has been developed, called initiator-FCIQMC (\iFCIQMCbracket)\cite{FCIQMCPaper2,EApaper,c2paper}. The determinant space is instantaneously divided into those determinants exceeding a population of $n_\text{add}$ walkers, termed initiator determinants, and those that do not. When considering a determinant \emph{whose current population is zero}, the sum in the second term of \refeq{FCI-eqs}, the term describing net flux of walkers onto that determinant, is taken to be only over initiator determinants.

\iFCIQMC has been shown to dramatically accelerate the convergence of FCIQMC with respect to walker number. In the large walker number limit, the \iFCIQMC tends to the FCIQMC algorithm, which itself converges rigorously to the FCI energy. In previous work, simulations with different walker numbers were performed to explicitly demonstrate convergence towards this limit by finding correlation energies over an increasing range of walker numbers\cite{c2paper}. In the present work we will show that this limit can be rigorously found from a single calculation.

\subsection{Previous Work on the HEG}
\label{prevwork}

In a previous study\cite{UEGPaper1}, \iFCIQMC was applied to the 54-electron HEG at $r_s$=0.5 and 1.0~a.u. to find energies for a range of $N_w$ and $M$. The $N_w \rightarrow\infty$ limit was found by direct evaluation using separate converged runs at different $N_w$ values, and the $M \rightarrow\infty$ limit was found by using a $1/M$ extrapolation. Finally, the resultant complete basis set exact energy compared favorably with diffusion Monte Carlo results (DMC)\cite{Rios}.

In these simulations, the walker number was grown in fixed shift mode under a set of parameters $S$, $\delta\tau$ and $n_\text{add}$ before being released into variable shift mode at a certain $N_w$. After being allowed to reach equilibrium, the finite walker \iFCIQMC energy ($E \left(N_w\right)$) was found from an imaginary time average of the projected energy (\refeq{projEeq}) which does not depend on $S$, since it is collected after equilibration in variable shift mode.

The form of the function $E \left(N_w\right)$ is however dependent on the two parameters $\delta\tau$ and $n_\text{add}$, which can be modified and optimized for efficiency. $E \left(N_w\right)$ is thought to vary the most with $n_\text{add}$, which must be kept at the same value for a set of simulations. The \iFCIQMC scheme tends towards the original FCIQMC method in the limit of $n_\text{add}=0$ or $N_w\rightarrow\infty$. A typical $n_\text{add}$ chosen is three, and we will analyze this choice later. In contrast, $E \left(N_w\right)$ is somewhat insensitive to the timestep $\delta\tau$, which is set to avoid too many spawning events causing unoccupied determinants to immediately form initiator determinants, since this can cause slow convergence. 

In the HEG, the plurality of matrix elements with the same magnitude means that slow convergence is observed if \emph{any} spawning events lead to immediate initiator formation, and as such, $\delta\tau$ is defined to be within the range,
\begin{equation}
\delta\tau < \frac {p_{\text{gen}} (\bfj|\bfi) n_{\text{add}}}{ |H_{\bfi\bfj}| }.
\label{tau_setting_eq}
\end{equation}
This limit is analytically computable for a given $N$ electrons and $M$ spin orbitals since generation probabilities in this case are uniform:
\begin{equation}
\left|\frac{p_{\text{gen}}}{H_{\bfi\bfj}}\right|_\text{min}=\frac{2}{N\left(N-1\right)}\frac{2}{M-N} \times \pi L.
\end{equation}
A $\delta\tau$ of approximately 90\% this maximum allowed value is used to maintain high acceptance ratios.

The finite walker \iFCIQMC energy, $E \left(N_w\right)$, obtained from a simulation has associated with it a systematic error due to the initiator approximation, which is rigorously removed in the limit of $N_w\rightarrow\infty$ when $E \left(N_w\right) \rightarrow E_\text{FCIQMC}$. The difference between $E \left(N_w\right)$ and this limit is termed initiator error.
%

\section{Analysis of \iFCIQMC and a fixed shift strategy for rapid convergence of initiator error}

\subsection{Division of initiator error and stochastic error}
\label{adaptation}

%



\begin{figure*}

  \subfloat[Estimate of $E \left(N_w\right)$ found by averaging $E_{\tau,i} \left(\tau \left(N_w\right) \right)$ over 120 seeds]{

\psfrag{KKK1}[l][l][1.0][0]{$E \left(N_w\right)$} 
\psfrag{BBB}[l][l][1.0][0]{$E_\text{FCIQMC}$} 
\psfrag{XXX}[b][b][1.2][0]{$N_w$} 
\psfrag{YYY}[b][b][1.2][0]{Correlation energy / $\text{E}_\text{h}$}

\includegraphics[width=0.45\textwidth]{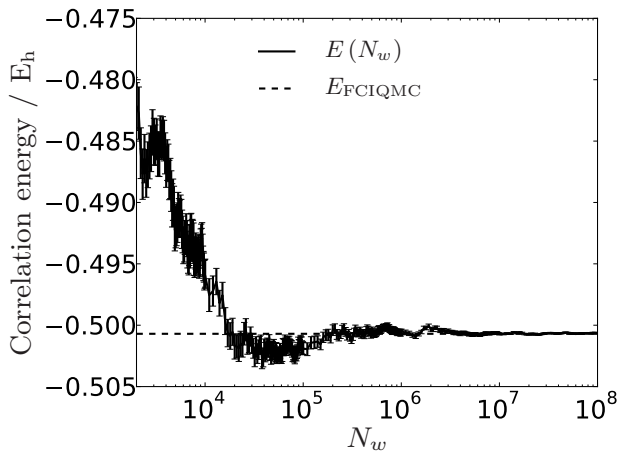}
  
  }\quad\quad
  \subfloat[Stochastic standard deviations for different numbers of seeds]{

\psfrag{KKK1}[l][l][1.2][0]{$N_\text{r}=120$} 
\psfrag{KKK2}[l][l][1.2][0]{$N_\text{r}=60$} 
\psfrag{KKK3}[l][l][1.2][0]{$N_\text{r}=30$} 
\psfrag{KKK4}[l][l][1.2][0]{$N_\text{r}=15$}

\psfrag{XXX}[t][t][1.2][0]{$N_w$} 
\psfrag{YYY}[t][t][1.2][0]{Stochastic standard deviation}
\psfrag{YYY2}[b][b][1.0][0]{($\sigma_\text{s}=\epsilon_\text{s} N_\text{r}^{1/2}$) / $\text{E}_\text{h}$ }

\includegraphics[width=0.45\textwidth]{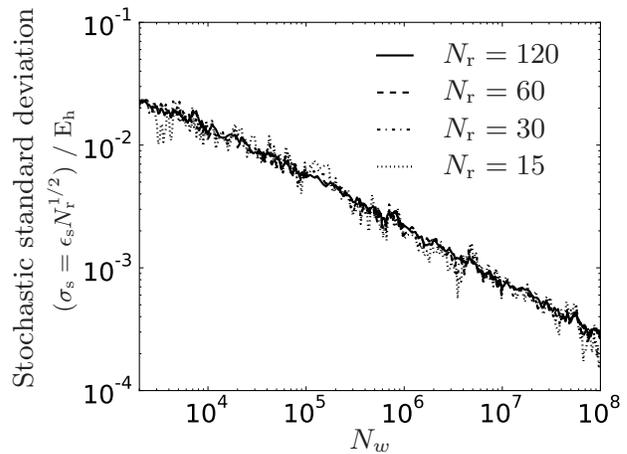}
\label{error_graph_2}

}

\caption{An example \iFCIQMC simulation performed on the $N=14$, $M=186$, $r_s=1.0$~a.u. electron gas with $n_\text{add}=3$. In the modified approach proposed here, fixed shift mode is used throughout the simulation (in this case $S=0.1 \text{E}_\text{h}$). The stochastic error is found by averaging over different pseudorandom number sequences, started from $N_r$ different seeds. This is assumed to be free from serial correlation. In the high $N_w$ limit, the FCI energy for the problem is recovered in common with the same limit in variable shift mode. In plot (b), the approach of finding the error from different seeds is justified. The stochastic standard deviation for various $N_r$ agree, implying the standard error decays as $N_r^{-\half}$.}

\label{normal_run}
\end{figure*}

We now describe a technique that allows us to better resolve the sources of error in the calculation and separate out stochastic error and initiator error. In so doing we will also place the technical observations of the previous study, described in \refsec{prevwork}, on a more rigorous footing.

In an \iFCIQMC run in fixed shift mode, as long as $S$ is higher than the correlation energy, the walker number grows exponentially, approximately as $e^{\tau\left(S-E_\text{corr}\right)}$. In this mode, the instantaneous projected energy (\refeq{projEeq}) will tend towards an increasingly stationary value in the large-$N_w$ limit, and settle onto the correlation energy for the problem. The ground state contribution to the wavefunction should always grow faster than any other excited state, and so regardless of the specific value of the shift parameter (which can be considered an energy offset in the Hamiltonian matrix) the ground-state should be recovered in the long-$\tau$ limit.

In a simulation where there is steady exponential growth, there is a one-to-one relationship between values of $\tau$ and $N_w$. The instantaneous projected energy for a single simulation can be written,
\begin{equation}
E_{\tau,i} \left( \tau \right) = \lim_{\tau\rightarrow\infty} \sum_{\bfj} \bra D_\bfj | H | D_{\bf 0} \ket \frac{ N_{\bfj} (\tau) }{ N_{\bf 0} (\tau) },
\end{equation}
and is therefore a well-behaved function of $N_w$. We assert that $E_{\tau,i} \left( \tau \right)$ written in terms of $N_w$ is therefore an approximation to the finite walker number \iFCIQMC energy $E \left(N_w\right)$. This is only rigorously true in the absence of any need for equilibration, when the simulation could be released into variable shift mode without a change in the average projected energy. This will be true as long as the growth in walker number is quasi-adiabatic. This estimate can be a poor representation of $E \left(N_w\right)$, since there is a large amount of stochastic error in each point. As with a normal simulation in variable shift mode each point is now serially correlated with the one before it. However, unlike in variable shift mode, the correlation time cannot be assumed to be constant because walker number is increasing. Therefore the blocking analysis due to Flyvbjerg and Petersen\cite{Flyvbjerg}, which is used to extract the correlation time, is no longer appropriate and we must investigate another method for estimating and minimizing the stochastic error.

Assuming points along $E \left(N_w\right)$ are unaffected by the starting point of the simulation, a straightforward way of finding the stochastic error would be to use several independent calculations, with different random number seeds. For $N_r$ seeds we can compute the instantaneous average,
\begin{equation}
E \left(N_w\right) \simeq E_{\tau} \left(N_w\right) = \frac{1}{N} \sum_i^{N_r} E_{\tau,i} \left(\tau \left(N_w\right) \right)
\end{equation}
where we have now used $\tau \left(N_w\right)$ to indicate that $\tau$ is a function of $N_w$. The stochastic error can be estimated from,
\begin{equation}
\epsilon_s  = \sqrt{ \frac{1}{N_r-1} \left( \sum_i^{N_r} \frac{ \left[E_i \left(\tau \left(N_w\right) \right)\right] ^2}{N_r}- \left(E_{\tau} \left(N_w\right)\right)^2\right)} .
\end{equation}
The specifics of each walker growth profile will mean that due to statistical fluctuations, identical values of $N_w$ can not be assumed for each simulation. Therefore, averages are taken in intervals for the closest values to a chosen set of $N_w$-values. 

\begin{figure}
\psfrag{KKK1}[l][l][0.9][0]{Cost of walker growth} 
\psfrag{XXX}[t][t][1.1][0]{Simulation time / corehours} 
\psfrag{YYY}[b][b][1.2][0]{$N_w$}
\includegraphics[width=0.45\textwidth]{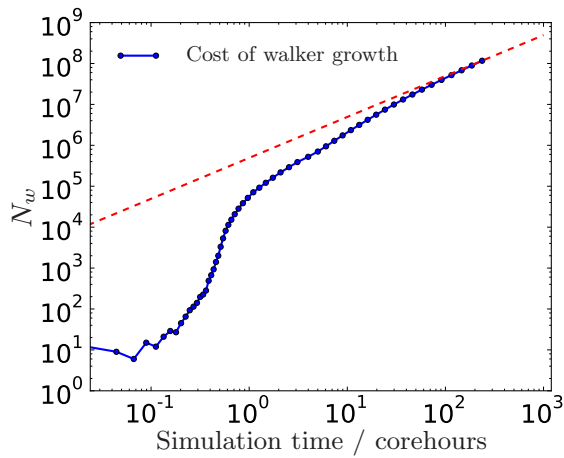}
\caption{Simulation time for $N=14$, $M=186$, $r_s=1.0$~a.u. and $n_\text{add}=3$ as a function of walker number. The formal scaling cost is $\mathcal{O}  \left[ N_w \text{Log} \left( N_w \right ) \right] $, but the scaling in the high walker number limit tends towards approximately $\mathcal{O} \left[ N_w \right]$ (shown by dashed, red line). The logarithmic scale is used to show the range of the scaling relationship.}
\label{time_graph}
\end{figure}

Figure \ref{normal_run} shows a simulation using this method, illustrating a way to approximate $E \left(N_w\right)$ without needing to obtain equilibrated energies from separate calculations at a set of $N_w$ values. To indicate that our estimate of the stochastic error is unbiased, it is also shown that the stochastic standard deviation, $\epsilon_s  N_{r}^{\frac{1}{2}}$, is preserved when the number of seeds is changed for a wide range of $N_w$ (\reffig{error_graph_2}). This verifies that the stochastic error falls off as $N_{r}^{-\frac{1}{2}}$ which is consistent with a good, uncorrelated error estimator. Also shown in this plot is the relationship between $N_w$ and the stochastic standard deviation, and we consider the trend to be of the form $N_w^{-\alpha}$ where $\alpha$ takes approximate values over the range shown in \reffig{error_graph_2} from $0.35 \lesssim \alpha \lesssim 0.47$. 

The trend observed in stochastic error is important to understand when discussing computational costs and scalings of the method. There are two ways in which the simulation can be modified to reduce the stochastic error. Either the the number of parallel runs can be increased or the number of walkers can be increased, resulting in polynomial $N_r^{-\half}$ and $N_w^{-\alpha}$ decay of error respectively. For a fixed-shift calculation, the simulation cost increase of walker growth is $\sim \mathcal{O} \left[ N_w \right]$ in the high $N_w$ limit (\reffig{time_graph}), with the memory cost also being linear in $N_w$. The cost in terms of memory and runtime of more $N_r$ also scales linearly. As separate copies of the program are run simultaneously, the parallelization over $N_r$ is perfectly linear. It is therefore more cost-effective to reduce the stochastic error for systems for $\alpha < 0.5$ using more parallel runs. 

However, increasing the number of walkers also decreases the initiator error. In practice, the rapidity with which initiator error decays still means that the $N_w \rightarrow \infty$ limit can be found by running the simulation at higher walker numbers until the energy does not change significantly with $N_w$. Thereafter, provided $\alpha < 0.5$, the stochastic error can be removed by using more random seeds. In the limit of converging onto the FCI wavefunction, $\alpha = 0.5$ due to increasingly fine discretization of $\Psi$, and these two methods of decreasing stochastic error become equivalent.

\subsection{The role of the shift parameter, $S$}

\begin{figure}

  \subfloat[$E_{\tau,S} \left( N_w , S \right)$ for different $S$ values]{
  
\psfrag{KKK1}[l][l][1.2][0]{$S=10.0$} 
\psfrag{KKK1b}[l][l][1.2][0]{$S=5.0$}
\psfrag{KKK2}[l][l][1.2][0]{$S=1.0$} 
\psfrag{KKK3}[l][l][1.2][0]{$S=0.1$} 
\psfrag{KKK4}[l][l][1.2][0]{$S=-1.0$} 
\psfrag{KKK5}[l][l][1.2][0]{Variable $S$} 

\psfrag{BBB}[l][l][1.0][0]{$E_{\iFCIQMC}$} 
\psfrag{XXX}[b][b][1.2][0]{$N_w$} 
\psfrag{YYY}[t][t][1.2][0]{Correlation energy / $\text{E}_\text{h}$}

\includegraphics[width=0.45\textwidth]{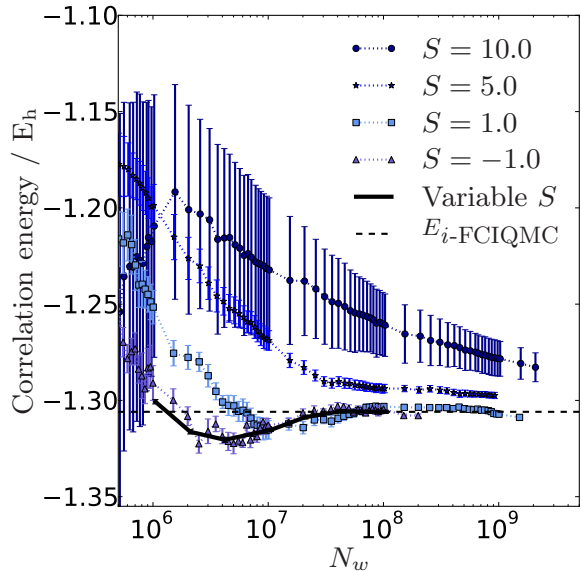}
}

  \subfloat[\iFCIQMC stochastic errors on $E_{\tau,S} \left( N_w , S \right)$ for different $S$ values]{
\psfrag{KKK1}[l][l][1.2][0]{$S=10.0$} 
\psfrag{KKK1b}[l][l][1.2][0]{$S=5.0$}
\psfrag{KKK2}[l][l][1.2][0]{$S=1.0$} 
\psfrag{KKK3}[l][l][1.2][0]{$S=0.1$} 
\psfrag{KKK4}[l][l][1.2][0]{$S=-1.0$} 
\psfrag{KKK5}[l][l][1.2][0]{$S$ varied} 

\psfrag{BBB}[l][l][1.0][0]{$E_{\iFCIQMC}$} 
\psfrag{XXX}[b][b][1.2][0]{$N_w$} 
\psfrag{YYY}[t][t][1.2][0]{Stochastic error / $\text{E}_\text{h}$}

\includegraphics[width=0.45\textwidth]{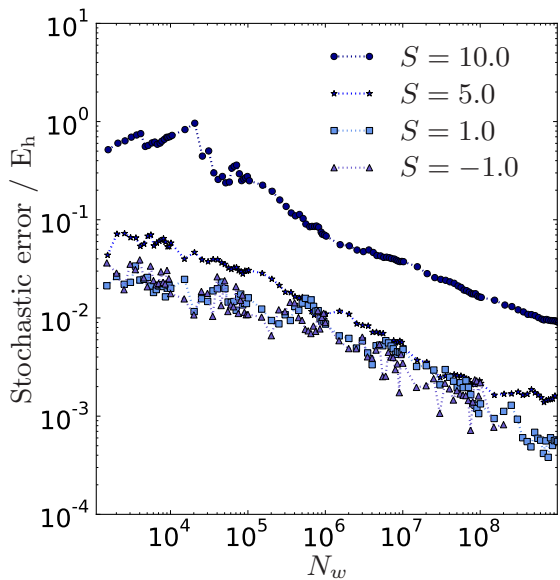}
}

\caption{\iFCIQMC runs at different values of the shift parameter, $S$, all performed using $N=54$, $M=186$, $r_s=1.0$, $n_\text{add}=3$ and $N_r=8$. These estimates of $E\left(N_w\right)$ tend towards the independent variable shift \iFCIQMC calculations as $S$ tends towards $E_\text{corr}$ (panel a). The relaxation towards this limit is observed to be exponentially fast in $\Delta=S-E_\text{corr}$. In the low $\Delta$ limit, the stochastic error does not change significantly between different values of $S$ (panel b).}
\label{shift_graph}
\end{figure}
\begin{figure}

\psfrag{KKK1}[l][l][1.2][0]{Rate of growth} 

\psfrag{BBB}[l][l][1.0][0]{$E_{\iFCIQMC}$} 
\psfrag{XXX}[t][t][1.0][0]{Rate of growth / $N_w$ (corehours)$^{-1}$ $\times 10^{-8}$} 
\psfrag{YYY}[t][t][1.2][0]{$S$ / $\text{E}_\text{h}$}

\includegraphics[width=0.45\textwidth]{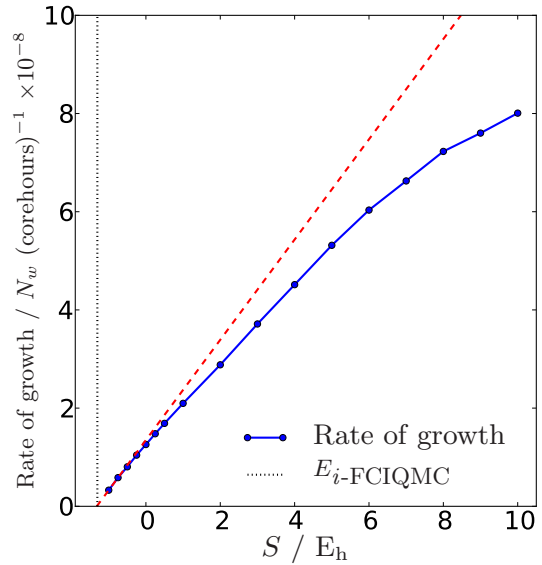}

\caption{\iFCIQMC runs in fixed-shift mode at different values of the shift parameter, $S$, all performed using $N=54$, $M=186$, $r_s=1.0$, $n_\text{add}=3$ and $N_r=1$. The rate of growth of walkers per corehour was measured from the linear, high $N_w$ limit (\reffig{time_graph}). Error bars were found on the fit on the order of 0.01\%. The speed of growth in the low $S$ limit grows linearly in $\Delta=S-E_\text{corr}$ from theoretical zero-rate growth at $S=E_\text{corr}$ (red dashed line). The next leading order term appears to be exponential, but this might be due to lack of convergence for the high-shift values (\reffig{shift_graph}). Nonetheless, linear growth rates are demonstrated for high $N_w$.}
\label{shift_vs_cputime}
\end{figure}
\begin{figure}

\psfrag{KKK1}[l][l][1.2][0]{Cost of growth} 
\psfrag{KKK2}[l][l][1.2][0]{Compound cost} 

\psfrag{BBB}[l][l][1.0][0]{$E_{\iFCIQMC}$} 
\psfrag{XXX}[t][t][1.2][0]{$\Delta$ / $\text{E}_\text{h}$} 
\psfrag{YYY}[t][t][1.2][0]{Relative computational cost to}
\psfrag{YYY2}[t][t][1.2][0]{converge / (arb. units)}

\includegraphics[width=0.45\textwidth]{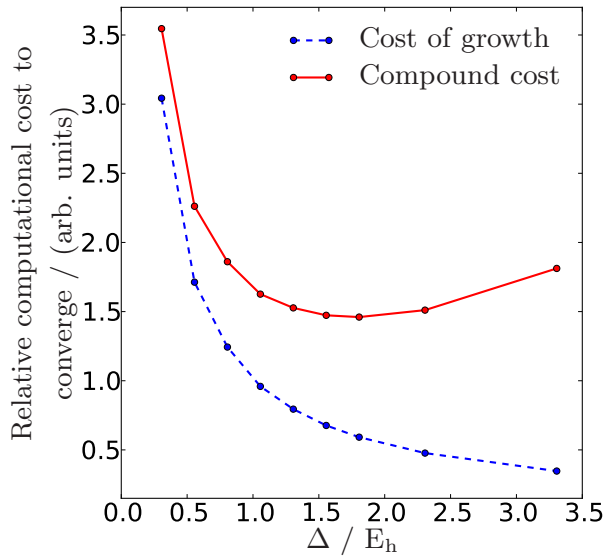}

\caption{\iFCIQMC runs in fixed-shift mode, at different values of the shift parameter, $S$, all performed using $N=54$, $M=186$, $r_s=1.0$, $n_\text{add}=3$ and $N_r=1$. The relative cost of converging to the large walker limit for a given shift value is estimated from a combination of the cost of growth from \reffig{shift_vs_cputime} (dashed blue line), and the higher $N_w$ needed to converge a given initiator error seen in \reffig{shift_graph} (which is approximately exponential in $\Delta$ for small $\Delta$). As such, there is a broad minimum in compound cost for a range of $\Delta$, which we expect to be highly system-dependent.}
\label{shift_effic}
\end{figure}

We now wish to compare fixed shift and variable shift calculations for efficiency, and as such we now frame our discussion in terms of whether $E\left(N_w\right)$ is best derived from fixed-shift or variable-shift calculations. A crucial difference that this introduces is that $E\left(N_w\right)$ estimated from a fixed-shift calculation is dependent on the $S$ that the simulation is fixed at. As such we will denote this $E_{\tau,S}\left(N_w,S\right)$, to make clear this dependence the energy now has on the choice of fixed shift.

Comparison between $E\left(N_w\right)$ and $E_{\tau,S}\left(N_w,S\right)$ for a variety of shift values shows empirically that as $S$ is reduced towards the correlation energy the estimate gets closer to the variable shift estimate for that given walker number. This is because, as $S$ is reduced in fixed shift mode, growth of the population is slower and a greater amount of equilibration can occur between each increase in walker number. In the limit that $S = E_\text{corr}$, the population never grows and should be able to equilibrate perfectly, and therefore become equivalent to variable shift mode in terms of the quality of the wavefunction generated at a given walker number. We can therefore make the equivalence $E\left(N_w\right)\equiv E_{\tau,S}\left(N_w,S=E_\text{corr}\right)$.

When choosing the shift for a simulation, there is a trade-off between lowering the shift, so that the convergence to the large $N_w$ energy is faster, and the runtime penalty this incurs from slow walker growth. The time taken to reach a certain walker number is found to scale linearly with $1/\Delta$, $\Delta=S-E_\text{corr}$, for low $\Delta$ (\reffig{shift_vs_cputime}). In contrast, the penalty for having too high $S$ is exponential in $\Delta$ (\reffig{shift_graph}). There is a minimum in cost as $S$ is increased at the cross-over between these two scaling relationships (\reffig{shift_effic}). 

The scaling of the growth rate for low $\Delta$ (\reffig{shift_vs_cputime}), can be expressed as:
\begin{equation}
\lim_{N_w \rightarrow \infty} \frac{N_w}{T} \propto \Delta
\end{equation}
where $T$ is the total simulation time. This is directly related to the high-$N_w$ walker growth, which proceeds as,
\begin{equation}
N_w = e^{\beta \Delta},
\end{equation}
where $\beta$ is a system-dependent constant, proportional to the total elapsed imaginary time. The simulation time is instantaneously proportional to $N_w$, so total simulation time can be written,
\begin{equation}
\begin{split}
T&=\int^{\beta} N_w \left( \beta^\prime \right) \, d \beta^\prime, \\
&=\int^{\beta}  e^{\beta^\prime \Delta} \, d \beta^\prime, \\
&= \frac{1}{\Delta} e^{\beta \Delta}.
\end{split}
\end{equation}
Therefore, in the high $N_w$ limit $\frac{N_w}{T} \propto \Delta$ as required to yield a computational cost scaling as $1/\Delta$.

\subsection{The initiator threshold parameter, $n_\text{add}$}

\begin{figure}

  \subfloat[$E_{\tau,S} \left( N_w , S \right)$ energies for different $n_\text{add}$ values]{

\psfrag{KKK1}[l][l][1.2][0]{$n_\text{add}=2$} 
\psfrag{KKK2}[l][l][1.2][0]{$n_\text{add}=3$} 
\psfrag{KKK3}[l][l][1.2][0]{$n_\text{add}=16$} 

\psfrag{XXX}[t][t][1.2][0]{$N_w^\star = N_w / n_\text{add}$} 
\psfrag{YYY}[t][t][1.2][0]{Correlation energy / $\text{E}_\text{h}$}

\includegraphics[width=0.45\textwidth]{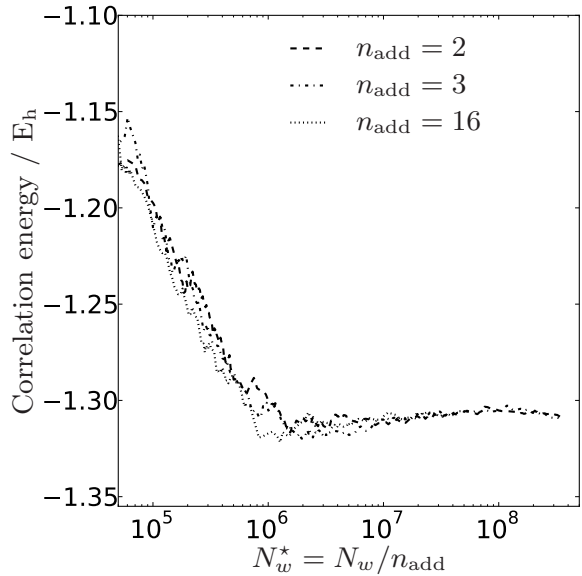}

}

  \subfloat[Stochastic errors on $E_{\tau,S} \left( N_w , S \right)$ for different $n_\text{add}$ values]{

\psfrag{KKK1}[l][l][1.2][0]{$n_\text{add}=2$} 
\psfrag{KKK2}[l][l][1.2][0]{$n_\text{add}=3$} 
\psfrag{KKK3}[l][l][1.2][0]{$n_\text{add}=16$} 

\psfrag{XXX}[t][t][1.2][0]{$N_w^\star = N_w / n_\text{add}$} 
\psfrag{YYY}[t][t][1.2][0]{Stochastic error / $\text{E}_\text{h}$}

\includegraphics[width=0.45\textwidth]{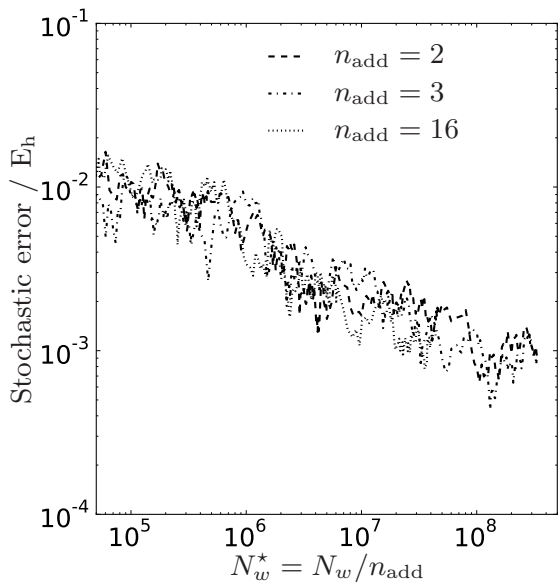}
\label{nadd_err_graph}
}

\caption{\iFCIQMC runs at different values of the $n_\text{add}$ parameter ($N=54$, $M=186$, $r_s=1.0$, $N_r=8$). To illustrate the apparent effect of changing $n_\text{add}$, the $N_w$ axis has been rescaled by dividing through by $n_\text{add}$, causing the lines to be overlaid.}
\label{nadd_graph}

\end{figure}

\begin{figure}[h]
\includegraphics[width=0.45\textwidth]{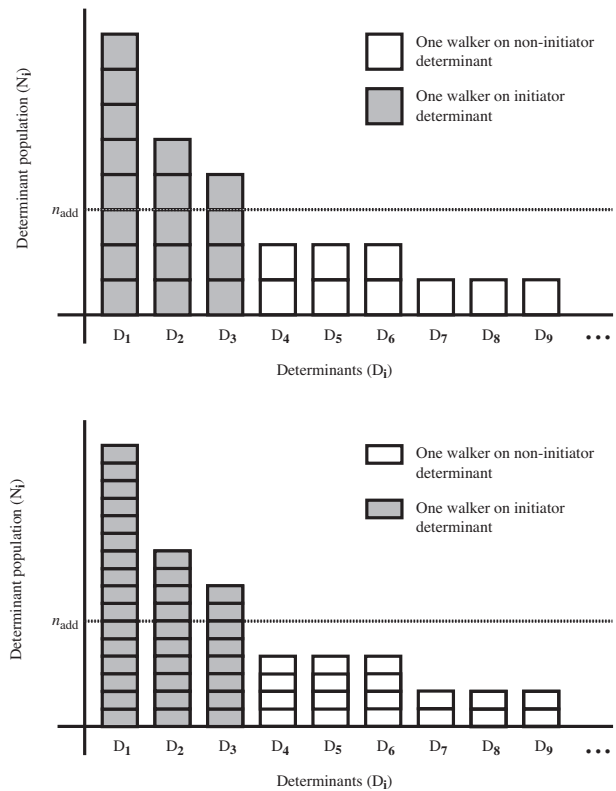}
\caption{Schematic diagram demonstrating how $n_\text{add}$ acts as a resolution parameter. In moving between the top and the bottom diagrams, $n_\text{add}$ has been doubled, but $N_w$ has also been doubled. This has no effect to the energy estimate, unless the ability to resolve low-weight determinants is important. In the high $N_w$ and $n_\text{add}$ limit this is unlikely to be the case.}
\label{ai_nadd_fig}
\end{figure}

The initiator threshold parameter $n_\text{add}$ determines the number of walkers above which an occupied site is considered an initiator and as such is an important parameter in \iFCIQMCbracket. In the limit of $N_w\rightarrow\infty$ all determinants become occupied and the algorithm returns to the original FCIQMC algorithm. The FCIQMC algorithm is also recovered in the limit of $n_\text{add}=0$ but this negates the computational advantages of the initiator approximation.

In principal, each value of $n_\text{add}$ should yield a different form of $E\left(N_w\right)$ since $n_\text{add}$ alters the effects of the dynamics in a non-trivial way. However, as \reffig{nadd_graph} shows, a much simpler relationship is observed. Although the relationship between energy and walker number is different for each value of $n_\text{add}$, this merely seems to rescale $N_w$ linearly. 

In this way $n_\text{add}$ can be seen to behave as a resolution parameter. Imagine comparing two simulations with $\{N_{w,A},n_{\text{add},A}\}$ and $\{N_{w,B},n_{\text{add},B}\}$. If $N_{w,B}=2 N_{w,A}$ and $n_{\text{add},B}=2 n_{\text{add},A}$, and assuming that there was a one-to-one mapping between determinant populations, $N_{\bfi,A}\rightarrow 2 N_{\bfi,B}$, the energy estimate at all paired values of $N_{w,I}^\star=N_w/n_\text{add}$ would be the same. This is demonstrated schematically in \reffig{ai_nadd_fig}. Although more walkers would normally lead to less stochastic error, the on-site and between-site flux would be rescaled with $n_\text{add}$ (\reffig{nadd_err_graph}). This analysis and explanation should only hold in the high $n_\text{add}$ and $N_w$ limit, and it just so happens that $n_\text{add}=2$ is high enough to display this behavior for the 54-electron $r_s=1.0$~a.u. HEG.


Finally, in these simulations, $\delta\tau$ was set according to \refeq{tau_setting_eq}. However, the form of $E_{\tau} \left(N_w\right)$ is independent of $\delta\tau$ for these systems provided the inequality \refeq{tau_setting_eq} is met. Since the run-time of a simulation is proportional to $\delta\tau$, this is simply maximized.

\subsection{Summary}

To summarize, in this section we have looked at a number of the simulation parameters for an \iFCIQMC calculation within the context of the problem posed by the HEG, and generalized where possible. From consideration of isolated `initiator' error arising from the finite number of walkers sampling the space, a new strategy was detailed whereby the calculation remains in fixed shift mode to reach high walker limits, while obviating the need for lengthy equilibration times in variable shift mode. These limits are required to demonstrate an effective elimination of this initiator error. This new simulation method was shown to be equivalent to previous schemes where separate converged calculations were performed for a variety of walker numbers. The cost of this strategy was critically analyzed in terms of speed of walker growth and decay of random errors, providing optimal values for the fixed shift, and initiator threshold parameter, $n_\text{add}$. In cases where this initiator error is challenging to converge, this strategy is presented as an appealing alternative. We mean this both in terms of effectively directing computational effort to ameliorate initiator and stochastic error, as well as provision of insight into the dynamics of the FCIQMC simulation in the space.

\section{Application to the HEG}

We now present an application of FCIQMC to the electron gas with the aim of producing results for the finite-basis 14-electron gas, and then also 
in the complete basis set limit with $N=14,38$ and $54$ electrons. 
Buoyed by various techniques to ameliorate the high scaling of quantum chemical methods, such as explicitly correlated 
basis sets\cite{Shiozaki2010,Yanai2012,Tew2012}, local approximations\cite{Usvyat2011,Scuseria2001,Schutz2007} and others\cite{Hirata2010,Kresse2011},
quantum chemists are beginning to tackle the problems presented in the solid-state. However, these efforts have generally excluded an examination
of the HEG. In producing high-accuracy literature benchmarks for the 14-electron gas, we 
hope that this will encourage the comparison of other techniques intended for application to solid state systems in the growing community looking 
to use quantum chemical techniques\cite{Kresse2009,Kresse2010,Pisani2007,Stoll2005,Paulus2012,Manby2010,Stoll2012}.

We now introduce an approximate extrapolation technique to efficiently calculate complete basis set limit estimates, with it in mind that these results can then be compared with DMC calculations. DMC calculations have been extremely successful in treating the HEG, with the most famous study being that of Ceperley and Alder\cite{CeperleyAlder1980}. Recent studies have tended to use the fixed-node approximation which intrinsically has an error associated with it, but that error is both thought to be small and unquantifiable\cite{Vignale}. These results are widely regarded as the best estimates of energies in the HEG over a range of densities. DMC has allowed for a large range of properties of the electron gas to be calculated, including phase diagrams\cite{CeperleyAlder1980,Drummond2009b}, the effective mass\cite{Drummond2009,Holzmann2009}, the renomalization factor\cite{Huotari}, spectral moments\cite{Needs} and the momentum distribution\cite{OB,OB2,Holzmann2011}. Unfortunately, since to the best knowledge of the authors no low-$N$ simulations exist in the literature for DMC, comparison between FCIQMC energies and DMC are left largely as an open question. It is intended that these results ultimately be used to compare between the initiator error of FCIQMC and the fixed-node error remaining in modern backflow DMC results.

\begin{figure}

\psfrag{KKK2}[l][l][1.2][0]{Single runs ($N_r=1$)} 
\psfrag{KKK3}[l][l][1.2][0]{$E_{\iFCIQMC} \left(N_r=20\right)$} 

\psfrag{X2}[lt][lt][1.2][0]{$1850^{-1}$} 
\psfrag{XL}[rt][rt][1.2][0]{$\infty^{-1}$} 
\psfrag{X3}[ct][ct][1.2][0]{$358^{-1}$} 
\psfrag{X4}[ct][ct][1.2][0]{$186^{-1}$}

\psfrag{XXX}[t][t][1.2][0]{$M^{-1}$} 
\psfrag{YYY}[t][t][1.2][0]{Correlation energy / $\text{E}_\text{h}$}

\includegraphics[width=0.45\textwidth]{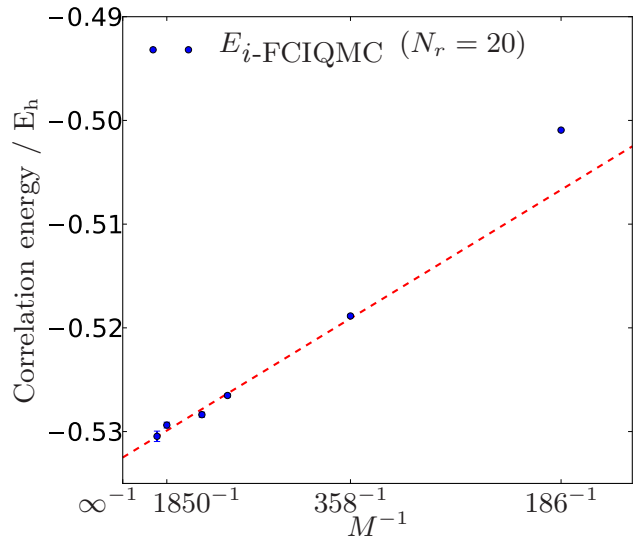}
\caption{\iFCIQMC results for $N=14$, $r_s=1.0$ showing a $1/M$ convergence to the CBS limit. These results were found to be converged with respect to initiator error at $N_w \sim 10^7-10^8$. Stochastic error bars are plotted but generally too small to be seen.}
\label{M_graph}

\end{figure}

\subsection{Basis set incompleteness error}
\label{BSSE}

Having dealt with the internal parameters that are important to the FCIQMC method, we now discuss the remaining parameter $M$, the size of our underlying one-electron basis, equal to twice the number of plane waves enclosed by a sphere centered at the origin of reciprocal space of radius $k_c$. Using this single parameter, the complete basis set (CBS) limit can be found by taking the limit of the correlation energy as $k_c\rightarrow\infty$.

The difference between the energy retrieved by a quantum chemical method in a finite basis, and the theoretical limit of this energy reached in a complete basis is termed basis set incompleteness error. Although a method may be a good approximation to exact results in principle, chemical accuracy is typically only achievable in the CBS limit. As such, there is much literature for how this limit is approached for molecular basis sets, and it has been shown both analytically and numerically that this limit is approached in $X$, the cardinal number of the basis set, as $1/X^3$ (Ref. \onlinecite{Kutzelnigg}).

In a separate paper yet to be published by the authors\cite{arxivextrappaper}, the CBS limit is shown to be approached as $1/M$ for plane-wave wavefunction methods. This is in agreement with the corresponding trend when using Gaussian expansions since the number of spin orbitals used scales as $X^3$. Figure \ref{M_graph} shows using this approach to obtain CBS limit energies for the $r_s=1.0$~a.u. 14-electron gas.

\subsection{Basis set scaling}

\begin{figure}

\psfrag{KKK1}[l][l][1.0][0]{Initiator error} 
\psfrag{KKK2}[l][l][1.0][0]{Stochastic error} 
\psfrag{XXX}[b][b][1.2][0]{$N_w$} 
\psfrag{YYY}[b][b][1.2][0]{Errors / $\text{E}_\text{h}$}

\includegraphics[width=0.45\textwidth]{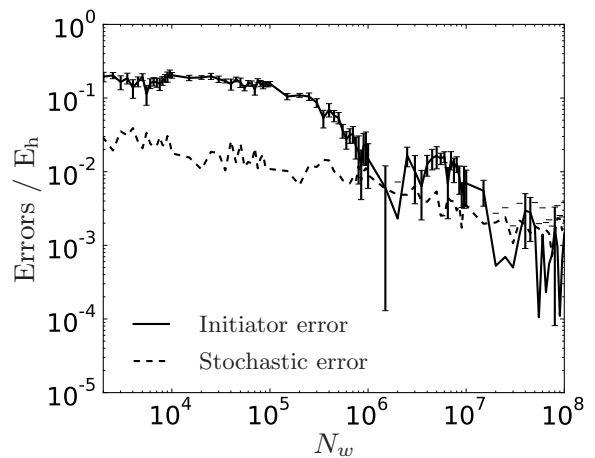}
\caption{Graph comparing initiator and stochastic error for $N=54$, $M=186$, $r_s=1.0$~a.u., $S=-0.1$~$\text{E}_\text{h}$ and $N_r=8$. The full curve of energies is shown in \reffig{shift_graph}.}
\label{noise_graph}

\end{figure}

\begin{figure}

  \subfloat[Initiator error does not change significantly with increase in basis set size in the high $M$ limit for this system, since all curves are simply a shift in energy from each other. Error bars are only shown for $M=114$ for clarity and dotted lines show asymptotic limits.]{

\psfrag{KKKKK1}[l][l][1.0][0]{$M=114$} 
\psfrag{KKKKK2}[l][l][1.0][0]{$M=186$} 
\psfrag{KKKKK3}[l][l][1.0][0]{$M=358$} 
\psfrag{KKKKK4}[l][l][1.0][0]{$M=778$} 
\psfrag{KKKKK5}[l][l][1.0][0]{$M=1850$} 
\psfrag{KKKKK6}[l][l][1.0][0]{$M=2368$} 

\psfrag{XXX}[t][t][1.2][0]{$N_\text{HF}$} 
\psfrag{YYY}[t][t][1.2][0]{Correlation energy / $\text{E}_\text{h}$}

\includegraphics[width=0.45\textwidth]{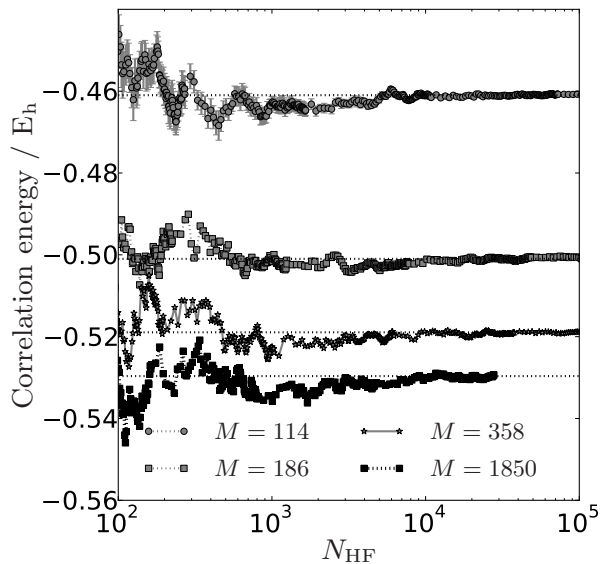}
\label{basis_set_graph}
}

  \subfloat[The stochastic error  as a function of a given population on the HF determinant ($N_\text{HF}$) does not change significantly with increase in basis set size in the high $M$ limit for this system.]{

\psfrag{KKKKK1}[l][l][1.0][0]{$M=114$} 
\psfrag{KKKKK2}[l][l][1.0][0]{$M=186$} 
\psfrag{KKKKK3}[l][l][1.0][0]{$M=358$} 
\psfrag{KKKKK4}[l][l][1.0][0]{$M=778$} 
\psfrag{KKKKK5}[l][l][1.0][0]{$M=1850$} 
\psfrag{KKKKK6}[l][l][1.0][0]{$M=2368$} 

\psfrag{XXX}[t][t][1.2][0]{$N_\text{HF}$} 
\psfrag{YYY}[t][t][1.2][0]{Stochastic error / $\text{E}_\text{h}$}

\includegraphics[width=0.45\textwidth]{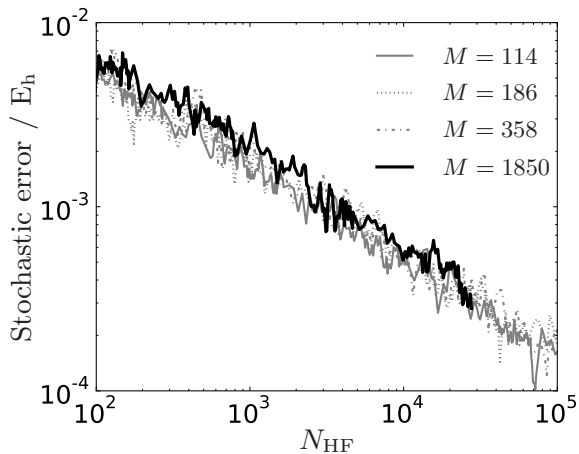}
\label{basis_set_graph_err_graph}

}

\caption{Initiator and stochastic error for \iFCIQMC runs at different basis set sizes, $N=14$, $r_s=1.0$, $N_r=8$ and $S=0.1$. 
}

\end{figure}

\begin{figure*}

  \subfloat[Change in fraction of the population at the Hartree-Fock determinant with walker population for a variety of basis set sizes]{

\psfrag{KKK}[r][r][1.2][0]{$M=$} 
\psfrag{KKK1}[l][l][1.2][0]{66} 
\psfrag{KKK2}[l][l][1.2][0]{114} 
\psfrag{KKK3}[l][l][1.2][0]{162} 
\psfrag{KKK4}[l][l][1.2][0]{246} 
\psfrag{KKK5}[l][l][1.2][0]{502} 
\psfrag{KKK6}[l][l][1.2][0]{1030} 
\psfrag{KKK7}[l][l][1.2][0]{1850} 
\psfrag{KKK8}[l][l][1.2][0]{2838} 

\psfrag{XXX}[t][t][1.2][0]{$N_w$} 
\psfrag{YYY}[t][t][1.2][0]{$N_\text{HF}/N_w$}

\includegraphics[width=0.45\textwidth]{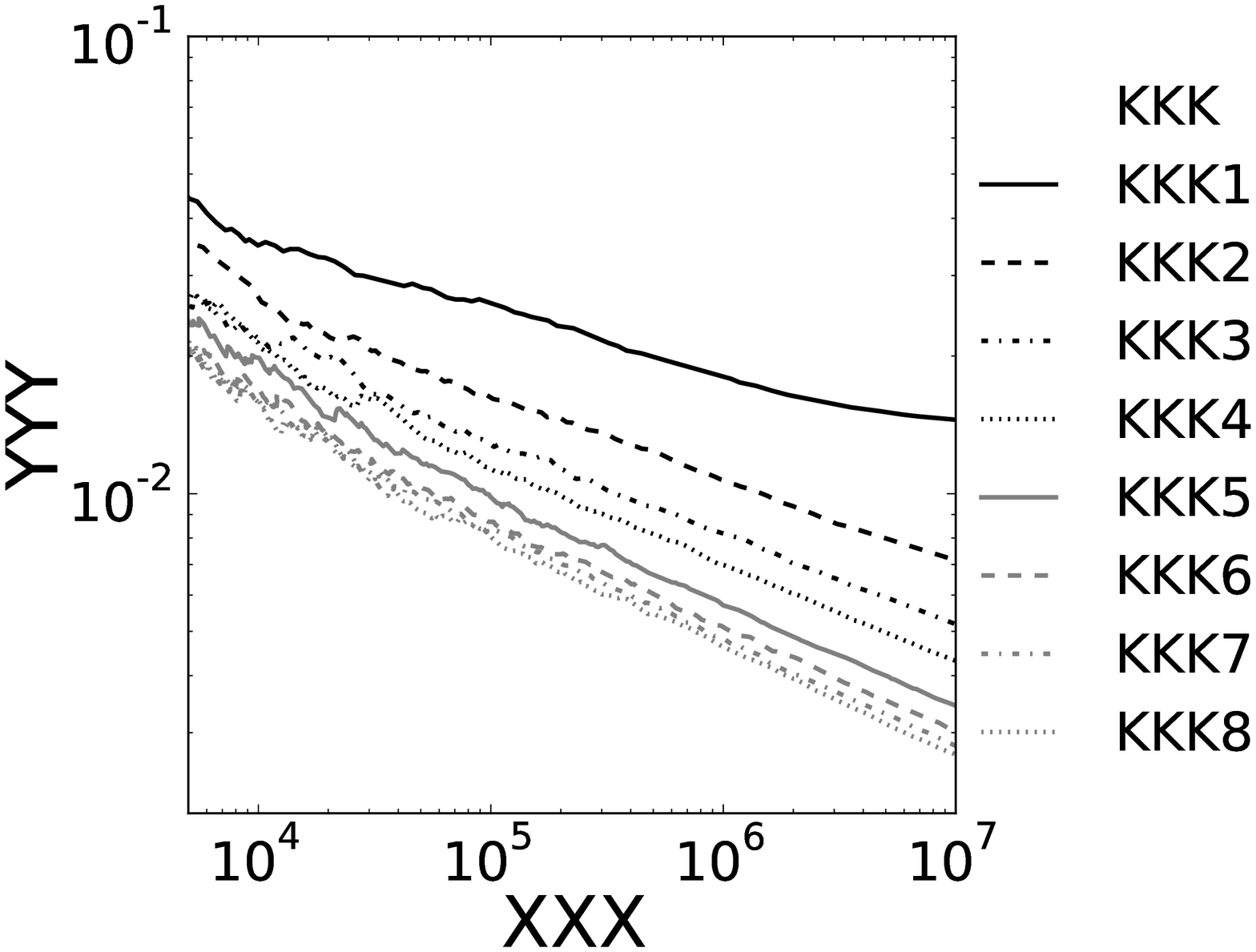}
\label{basis_set_graph_single_runs_2}

}
  \subfloat[The ratio of the population at the Hartree-Fock determinant to the total population of doubly excited determinants changes with basis set size but not with $N_w$ at convergence.]{

\psfrag{KKK}[r][r][1.2][0]{$M=$} 
\psfrag{KKK1}[l][l][1.2][0]{66} 
\psfrag{KKK2}[l][l][1.2][0]{114} 
\psfrag{KKK3}[l][l][1.2][0]{162} 
\psfrag{KKK4}[l][l][1.2][0]{246} 
\psfrag{KKK5}[l][l][1.2][0]{502} 
\psfrag{KKK6}[l][l][1.2][0]{1030} 
\psfrag{KKK7}[l][l][1.2][0]{1850} 
\psfrag{KKK8}[l][l][1.2][0]{2838} 

\psfrag{XXX}[t][t][1.2][0]{$N_w$} 
\psfrag{YYY}[t][t][1.2][0]{$N_\text{HF}/N_\text{doubs}$}

\includegraphics[width=0.45\textwidth]{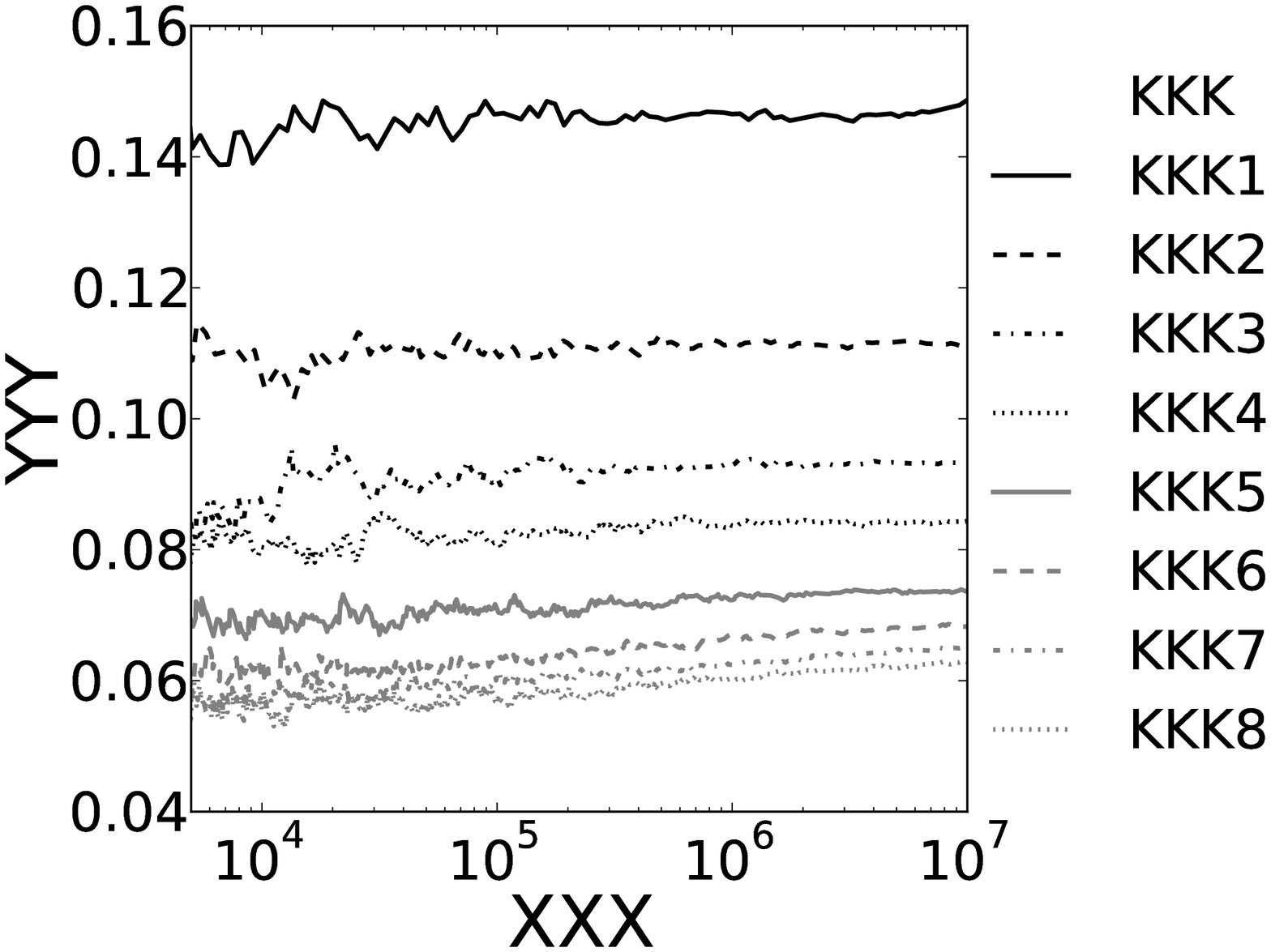}

\label{basis_set_graph_single_runs_1}
}
\caption{Trends in the \iFCIQMC wavefunction for $N=14$, $r_s=1.0$~a.u.} 
\end{figure*}

\begin{figure}

\psfrag{KKK2}[l][l][1.2][0]{$N_\text{HF}/N_w$ at $N_w=10^7$} 

\psfrag{XXX}[t][t][1.2][0]{$M^{-1}$} 
\psfrag{YYY}[t][t][1.2][0]{$N_\text{HF}/N_w$}

\includegraphics[width=0.45\textwidth]{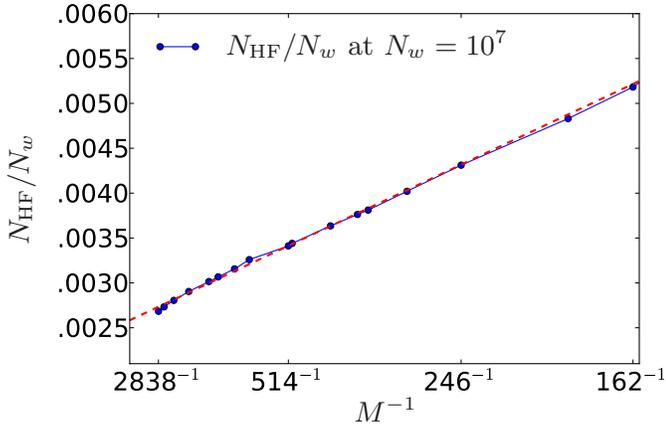}
\caption{When the initiator error is converged, for the $N=14$, $r_s=1.0$~a.u., the fraction of the walker population that is at the Hartree-Fock determinant falls as $1/M$ for large basis set sizes} 
\label{basis_set_graph_all_runs}

\end{figure}

An important aspect of any new method is the computational scaling of this method with parameters of the systems that are being studied. One such parameter of interest is basis set size, $M$. Since FCIQMC is a relatively new method, very little scaling work has been considered. Here, we present an initial analysis of basis set scaling as applied to the weakly correlated $N=14$, $r_s=1.0$~a.u. electron gas. 

Since FCIQMC is a stochastic method, the factors affecting the time required for the simulation can be crudely broken down into three considerations: 
\begin{enumerate}
\item The number of walkers required to converge the calculation. This is primarily to eliminate initiator error, but number of walkers also aids in convergence of stochastic error. Initiator error estimates are very difficult to quantify due to, in principle, the $N_w\rightarrow\infty$ limit needing to be reached for comparison. Moreover, since the CBS estimate comes with a stochastic error, as do other points along the initiator error graph, the initiator error becomes rapidly lost in stochastic noise (\reffig{noise_graph}). However, initiator error seems to decay very rapidly for systems studied and graphs of $E \left(N_w\right)$ with characteristic decays can be compared.
\item The simulation time taken to grow the number of walkers. Although this is $\mathcal{O}  \left[ N_w \text{Log} \left( N_w \right ) \right] $ in the current implementation, the difference between this and a linear scaling is very small in the high walker limit. Moreover, the algorithm could be theoretically optimized for linear-scaling growth at all $N_w$ but this is thought to have a more costly prefactor. As such, this will be generally discussed as linear-scaling in $N_w$.
\item The simulation time taken to reduce the stochastic error. This is best achieved from the point of view of the present work by increasing the number of seeds for the simulation $N_r$. 
\end{enumerate}

The number of walkers on the Hartree-Fock determinant required to converge the calculation, from the point of view of initiator error, is startlingly invariant with respect to $M$ for the system studied here (\reffig{basis_set_graph}). Initiator error plots appear essentially identical in the high $M$ limit. In addition, the stochastic error decays at the same rate for each $M$ with respect to the population on the Hartree-Fock determinant ($N_{\text{HF}}$), as shown in \reffig{basis_set_graph_err_graph}. However, the simulation time taken to reach the number of walkers required, however, increases as $\mathcal{O} \left[M\right]$ because the connectivity of the space grows as this factor, and therefore $\delta\tau$ is reduced to maintain the same quality of sampling. This factor seems to be the leading order scaling in $M$.  

Although $N_{\text{HF}}$ would be expected to grow proportionally to $N_w$ at convergence, it only grows as $\left(N_w\right)^\gamma$ for $\gamma < 1$ for typical HEG simulations. This can be seen from the ratio of the population on the Hartree-Fock determinant to the total walker population, which should be constant for $\gamma =1$ (\reffig{basis_set_graph_single_runs_2}). This trend actually predicts $\gamma \simeq 0.76$ since a number of walkers at any time reside on low amplitude determinants, which are generally very large in number, pushing $N_\text{HF}/N_w$ down. Although this would indicate a lack of convergence, the ratio of $N_\text{HF}$ to the population at the double excited determinants, $N_\text{doubs}$, is approximately constant within stochastic fluctuations when a calculation has converged (\reffig{basis_set_graph_single_runs_1}). These are the only contributing determinants to the projected energy, the energy estimate used in this study, and so it makes sense that when there is a stationary distribution of walkers across these determinants, the simulation would be converged. 

Returning to the question of scaling with $M$, the fraction $\frac{N_\text{HF}}{N_w}$ is shown in \reffig{basis_set_graph_all_runs}, and generally behaves as $1/M$, for high walker numbers. The total number of walkers taken to reach a given `target' population at the HF determinant is therefore $N_w=\frac{M}{A+BM}$, where $A$ and $B$ are constants. The leading order contribution to this in the high $M$ limit is $\mathcal{O} \left[ 1 \right]$. This, then, should be multiplied by the cost of walker growth, $\mathcal{O} \left[M \right]$ due to the required decrease in $\delta\tau$, to give $\mathcal{O} \left[M \right]$, or linear scaling, overall. This linear scaling is only physically realistic if we consider that the basis functions in the high-energy parts of the space are completely decoupled from one another. This could well be very reasonable for the high $M$ limit of a weakly correlated system.

The final consideration to make is to comment that we have only observed this behavior for the relatively small system of $N=14$, and the relatively weakly-correlated $r_s=1.0$~a.u. How transferrable are these findings? It is probable that the observation of a constant initiator error with $M$ will not hold for larger, or more correlated, systems. Indeed it has already been shown that for the $N=54$ electron gas that there is a significant change of initiator error with basis set size\cite{UEGPaper1}, although it is likely that the high basis set limit is reached much more quickly for $N=14$. Notwithstanding this, it is hard to think that computational effort would scale in any way other than exponentially in $M$, since the size of the space to be sampled is growing as $\mathcal{O}\left[M ! \right]$, but that the prefactor might be low enough that this is never observed within the desired random error.

In spite of this somewhat surprising scaling relation, the 14 electron problem is not a trivial one for quantum chemical methods and upon entering the linear scaling regieme, the remaining basis set incompleteness error is still of the order $\sim$0.01~$\text{E}_\text{h}$ (\reffig{basis_set_graph}). As such the finding that the high $M$ limit of this system can be captured as $\mathcal{O}\left[M\right]$ is important. The apparent lack of growth of initiator error on increasing $M$ shows that the sparsity with which the space to be sampled does not grow significantly with $M$, which may well apply to other systems.

\subsection{Comparison of electron densities ($r_s$)}

\begin{table}[h]
\begin{tabular*}{0.45\textwidth}{@{\extracolsep{\fill}} | c | c | c | c | c | }
 \hline
  & \multicolumn{4}{c |}{$r_s$ (a.u.)} \\
M & \multicolumn{1}{c}{0.5}  & \multicolumn{1}{c}{1.0}  & \multicolumn{1}{c}{2.0}  & 5.0   \\
  \hline      
  \hline  
114 & 	-0.5169(1) & 	-0.46111(9) & 	-0.3842(2) & -0.2645(3) \\ 
186 & 	-0.5589(1) & 	-0.50093(9) & 	-0.4207(3) & -0.2928(4) \\ 
358 & 	-0.5797(3) & 	-0.5189(1) & 	-0.4355(4) & -0.3017(7) \\ 
778 & 	-0.5893(3) & 	-0.5265(2) &	-0.4410(5) & -0.304(1) \\ 
1850 & 	-0.5936(3) & 	-0.5294(3) & 	-0.4431(5) &  \\ 
2368 & 	-0.5939(4) & 	-0.5305(5) & 	-0.4430(7) & \\ 
\hline
$\infty$ & 	-0.5969(3) & 	-0.5325(4) & 	-0.4447(4) & -0.306(1) \\ 
  \hline  
\end{tabular*}
  \caption{\iFCIQMC correlation energies for $N=14$. The number in brackets corresponds to the stochastic error in the preceeding digit. The $M=\infty$ result is based on extrapolations shown in \reffig{M_graph} from which its error estimate derives. }
  \label{onlytable}
\end{table}

\begin{figure}

\psfrag{KKK1}[c][c][1.2][0]{$r_s=0.5$} 
\psfrag{KKK2}[c][c][1.2][0]{$r_s=1.0$} 
\psfrag{KKK3}[c][c][1.2][0]{$r_s=2.0$} 
\psfrag{KKK4}[c][c][1.2][0]{$r_s=5.0$} 
\psfrag{BBB}[c][c][1.2][0]{$\; N_w\rightarrow\infty$} 

\psfrag{XXX}[t][t][1.2][0]{$N_w$} 
\psfrag{YYY}[t][t][1.2][0]{Correlation energy / $\text{E}_\text{h}$}

\includegraphics[width=0.45\textwidth]{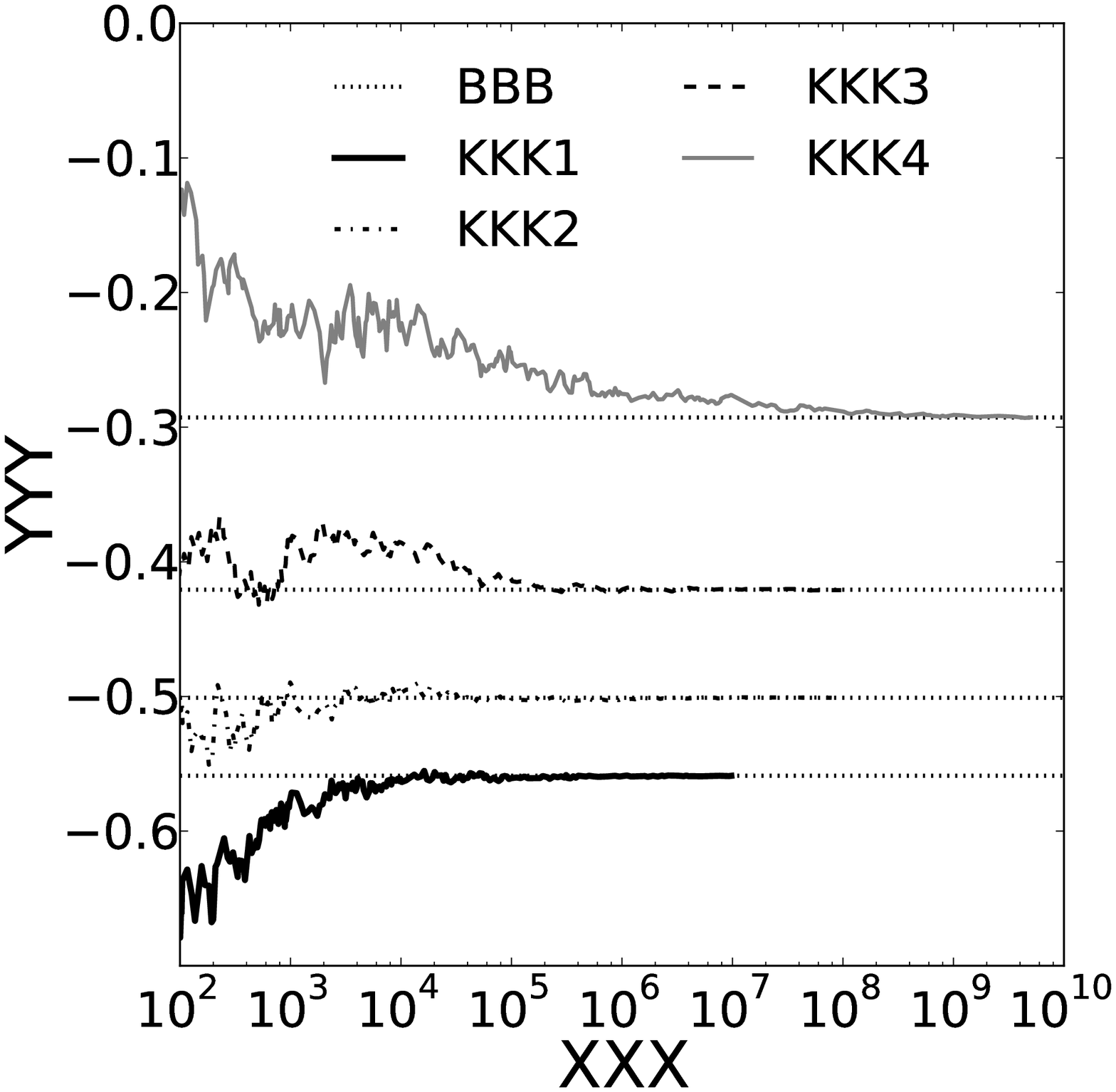}
\caption{Graphs of correlation energy retrieved with respect to walker number with different $r_s$, $M=186$. The $N_w\rightarrow\infty$ limit corresponds to the correlation energy for each of the systems. Convergence to this limit is slower in $N_w$ for larger $r_s$-values.}
\label{rs_scaling}
\end{figure}

\begin{figure}

\psfrag{KKK1}[l][l][1.2][0]{$r_s=0.5$} 
\psfrag{KKK2}[l][l][1.2][0]{$r_s=1.0$} 
\psfrag{KKK3}[l][l][1.2][0]{$r_s=2.0$} 
\psfrag{KKK4}[l][l][1.2][0]{$r_s=5.0$} 
\psfrag{BBB}[l][l][1.2][0]{CBS extrapolations} 

\psfrag{XY}[t][t][1.2][0]{$\infty^{-1}$} 
\psfrag{XXX}[c][c][1.2][0]{$|$} 

\psfrag{XXXB}[t][t][1.2][0]{$M^{-1}$} 
\psfrag{YYY}[t][t][1.2][0]{Basis set incompleteness error / $\text{E}_\text{h}$}

\psfrag{X2}[lt][lt][1.2][0]{$1850^{-1}$} 
\psfrag{X3}[ct][ct][1.2][0]{$358^{-1}$} 
\psfrag{X4}[ct][ct][1.2][0]{$186^{-1}$} 

\includegraphics[width=0.45\textwidth]{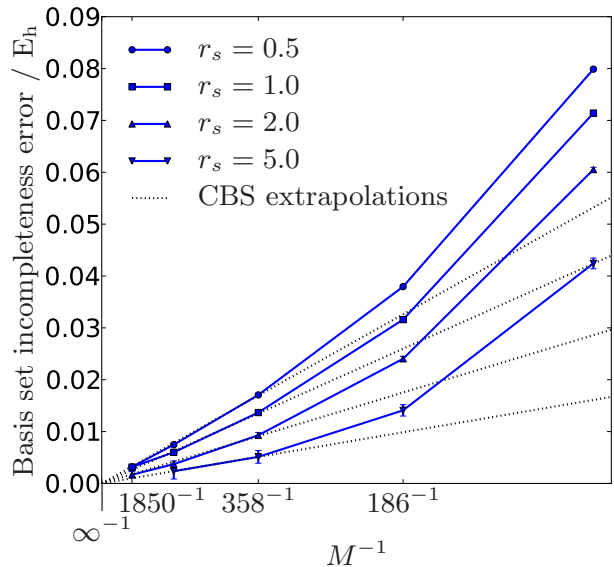}
\caption{Basis set incompleteness error for $0.5 \leq r_s \leq 5.0$~a.u. for a variety of basis set sizes. Convergence is of $1/M$ form (dotted black lines) in the limit of $M\rightarrow\infty$, and the rate of convergence to this limit does not appear to change as $r_s$ is raised. This is partly because the magnitude of the basis set incompleteness error decreases with increasing $r_s$, and therefore an estimate of the complete basis set result is less sensitive to the precise form and behavior of the extrapolation.}
\label{rs_basis_graph}
\end{figure}

FCIQMC energies obtained for $r_s=0.5-5.0$~a.u. are given in \reftab{onlytable}, which we present as new small-system benchmarks. To our knowledge, systems of this small an electron number but with nonetheless vast Hilbert spaces, have not been studied to date and as such we have no values to compare to. In a previous study, however, we showed that FCIQMC results are highly competitive with DMC results in the complete basis limit\cite{UEGPaper1}.

As $r_s$ is raised, the difficulty of the problem for \iFCIQMC rises sharply, which we can see from the energy retrieved against walker number in \reffig{rs_scaling}. Whilst the $r_s=0.5$ calculation for the basis set shown is converged at $10^4$ walkers, the $r_s=5.0$ calculation takes $10^9$ walkers to converge. We can attribute this to the lowering of the sparsity of this representation of the wavefunction due to stronger correlation effects at larger $r_s$. We anticipate that the $r_s$ parameter would behave similarly in a conventional FCIQMC calculation to the Hubbard $U$ parameter, whose effect on the sign problem in FCIQMC has been analysed in detail\cite{Spencer}.
The extrapolation to the complete basis set limit for these densities is shown in \reffig{rs_basis_graph}, and indicate that the onset of the $1/M$ scaling regime is relatively insensitive to this density. This is also demonstrated later in \reffig{rs_splitproje}.

\subsection{Using a single point extrapolation of the projected energy to achieve complete basis set estimates}

\begin{figure}

\psfrag{KKK2}[l][l][1.0][0]{$M=1850$ single point extrapolation} 
\psfrag{KKK3}[l][l][1.0][0]{$E_{\iFCIQMC}$} 

\psfrag{XG2}[lt][lt][1.2][0]{$358^{-1}$} 
\psfrag{XGL}[rt][rt][1.2][0]{$\infty^{-1}$} 
\psfrag{XG3}[ct][ct][1.2][0]{$114^{-1}$} 
\psfrag{XG4}[ct][ct][1.2][0]{$54^{-1}$} 
\psfrag{XG5}[ct][ct][1.2][0]{$38^{-1}$}

\psfrag{XXX}[t][t][1.0][0]{$M^{-1}$ or $M^{\prime -1}$} 
\psfrag{YYY}[t][t][1.2][0]{Correlation energy / $\text{E}_\text{h}$}

\includegraphics[width=0.45\textwidth]{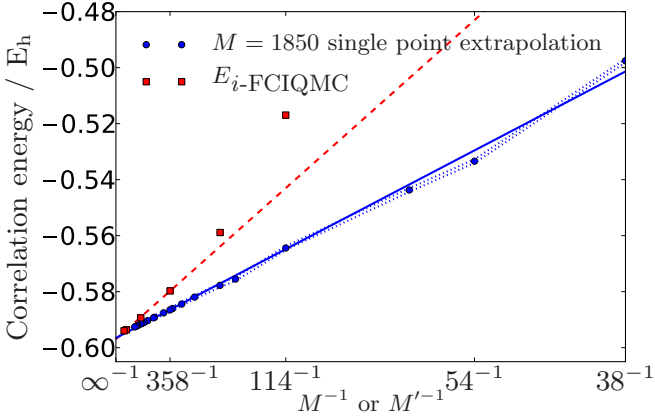}
\label{proje_g_graph}

\caption{\iFCIQMC results for $N=14$, $r_s=0.5$ showing a $1/M$ or $1/M^\prime$ convergence to the CBS limit for two schemes. In the conventional scheme, calculations are run at different basis set sizes and extrapolated to the CBS limit (\refsec{BSSE}). However, it is also possible to find a CBS estimate from single point extrapolation (solid blue line) of the projected energy, where the points along this line derive from a single calculation at an overall basis set size $M$=1850 (see text). As such, they share a single stochastic error bar (dotted blue line). These results were found to be converged with respect to initiator error at $N_w \sim 10^6-10^7$.}

\end{figure}

\begin{figure}

\psfrag{KKK1}[l][l][1.0][0]{$E_{\iFCIQMC}$} 
\psfrag{KKK2}[l][l][1.0][0]{Single point extrapolation} 
\psfrag{KKK3}[l][l][1.0][0]{Direct extrapolation} 
\psfrag{BBB}[l][l][1.0][0]{CBS estimate} 

\psfrag{X2}[ct][ct][1.1][0]{$778^{-1}$} 
\psfrag{XL}[ct][ct][1.1][0]{$\infty^{-1}$} 
\psfrag{X3}[ct][ct][1.1][0]{$358^{-1}$} 
\psfrag{X4}[ct][ct][1.1][0]{$186^{-1}$} 
\psfrag{X5}[ct][ct][1.1][0]{$114^{-1}$}

\psfrag{XXX}[t][t][1.0][0]{$M^{-1}$ (direct and SPE)} 
\psfrag{YYY}[t][t][1.2][0]{Correlation energy / $\text{E}_\text{h}$}

\includegraphics[width=0.45\textwidth]{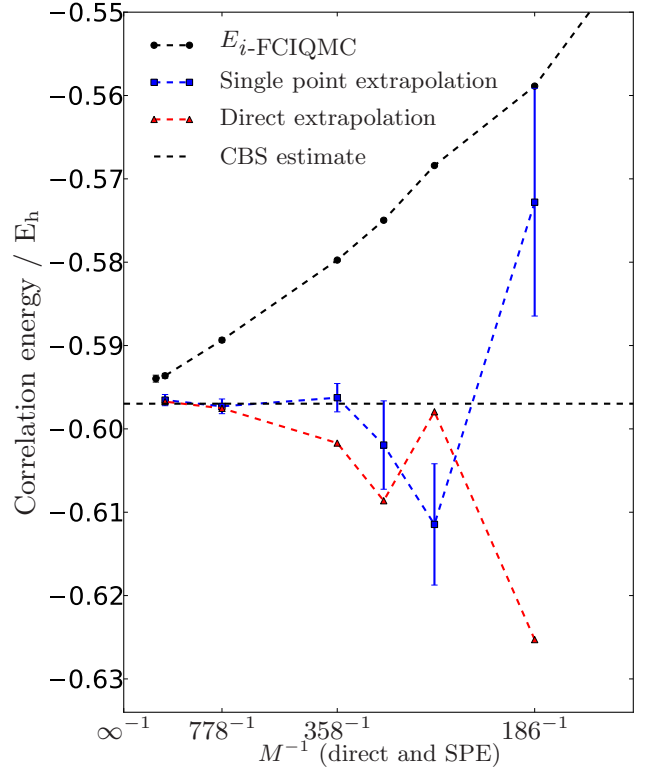}
\caption{Comparison between results obtained by conventional \iFCIQMC calculations with CBS estimates from either normal extrapolation, described in \refsec{BSSE}, or single point extrapolations (see text). At each value of $M$ an \iFCIQMC simulation was run to yield an energy. This energy was then either extrapolated directly or a single-point extrapolation from this value of $M$ was used to estimate the CBS limit using the masking function described in the text.}
\label{proje_graph}

\end{figure}

\begin{figure}

\psfrag{KKK1}[l][l][1.2][0]{$r_s=0.5$} 
\psfrag{KKK2}[l][l][1.2][0]{$r_s=1.0$} 
\psfrag{KKK3}[l][l][1.2][0]{$r_s=2.0$} 
\psfrag{KKK4}[l][l][1.2][0]{$r_s=5.0$} 
\psfrag{BBB}[l][l][1.2][0]{CBS extrapolations} 

\psfrag{XY}[t][t][1.2][0]{$\infty^{-1}$} 
\psfrag{XXX}[c][c][1.2][0]{$|$} 

\psfrag{XXXB}[t][t][1.2][0]{$M^{\prime -1}$} 
\psfrag{YYY}[t][t][1.2][0]{Basis set incompleteness error / $\text{E}_\text{h}$}

\psfrag{X2}[lt][lt][1.2][0]{$1030^{-1}$} 
\psfrag{X3}[ct][ct][1.2][0]{$162^{-1}$} 
\psfrag{X4}[ct][ct][1.2][0]{$114^{-1}$} 
\psfrag{X5}[ct][ct][1.2][0]{$66^{-1}$} 
\psfrag{X6}[ct][ct][1.2][0]{$54^{-1}$} 

\includegraphics[width=0.45\textwidth]{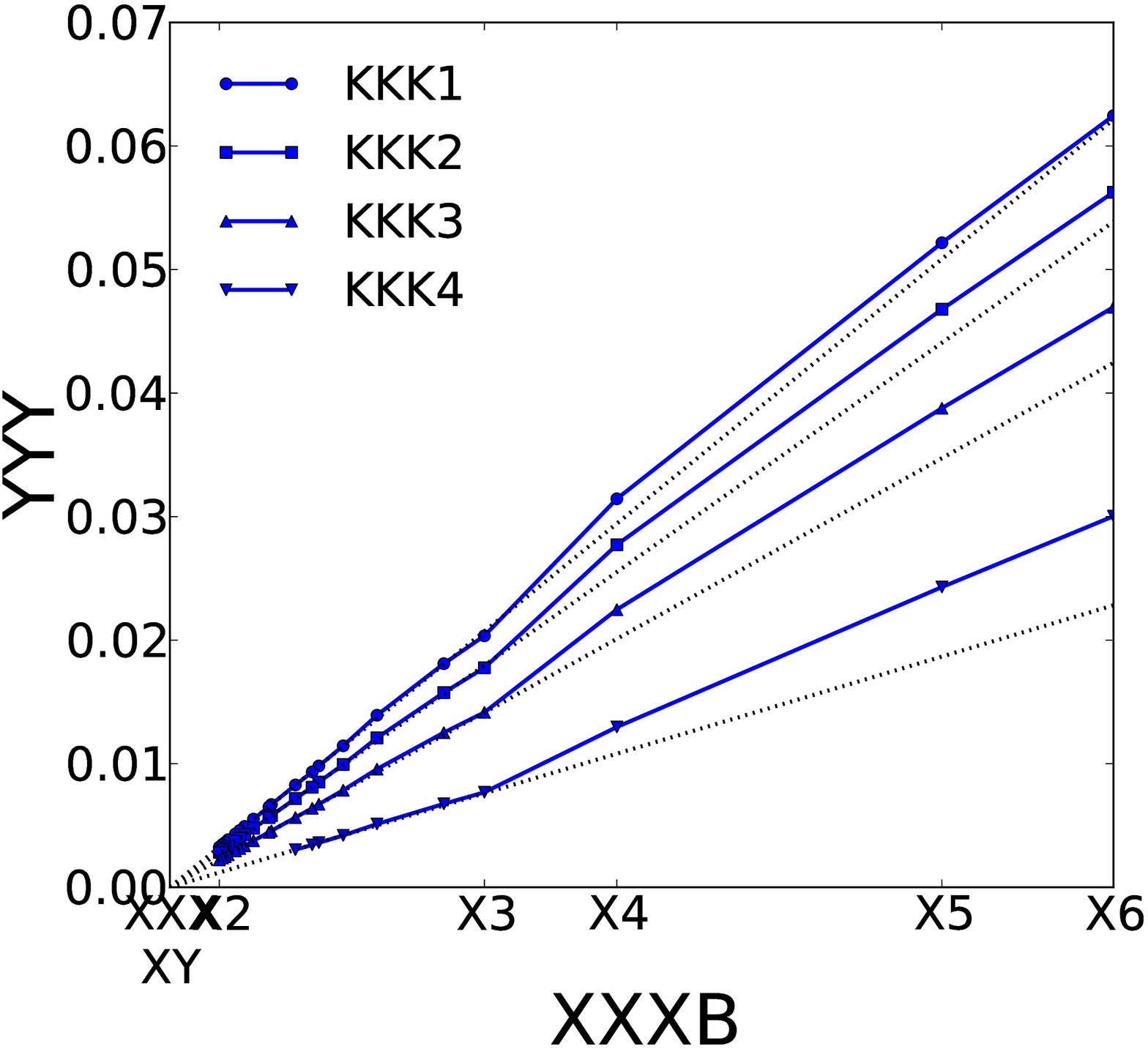}
\caption{Basis set incompleteness error for $0.5 \leq r_s \leq 5.0$ for a variety of effective basis set sizes. In contrast to \reffig{rs_basis_graph}, the $M^\prime$ label here refers to an effective basis set size derived during single point extrapolation of the projected energy. Convergence is of $1/M^\prime$ form (dotted black lines) in the limit of $M^\prime\rightarrow\infty$ and this limit does not appear to be approached more slowly in $M^\prime$ as $r_s$ is raised, although the curvature is more pronounced at high $r_s$. The basis set size $M$ for these calculations is given in \reftab{table2}.}
\label{rs_splitproje}
\end{figure}

In a separate study of basis set convergence in plane wave wavefunction methods by the authors, yet to be published, it was shown that it is possible to use a single large basis set calculation to yield an estimate of the CBS correlation energy for the HEG. This is achieved by dividing the contributions to the energy from a single large basis set calculation into regions of momentum space, producing smaller, effective basis sets from which the CBS limit can be estimated by extrapolation. 

Starting from the formulation of the FCIQMC correlation energy that we are using, the projected energy (defined in \refeq{projEeq}),
\begin{equation}
E_{\text{proj}} = \sum_{\bfj} \bra D_\bfj | H | D_{\bf 0} \ket \frac{ c_{\bfj}}{ c_{\bf 0}},
\end{equation}
where $\bfj$ refers to double excitations of the Hartree-Fock determinant. We can divide this into individual contributions from sets of four $k$-points, which uniquely define the four one-electron states for each double excitation $ij\rightarrow ab$, where $ij$ are occupied in the Hartree-Fock determinant and $ab$ are unoccupied,
\begin{equation}
E_{\text{corr}} =
\sum_{ij}^{\text{occ}} \sum_{ab}^{\text{virt}} \chi_{\bfk_i \bfk_j}^{\bfk_a \bfk_b}.
\end{equation}
with,
\begin{equation}
\chi_{\bfk_i \bfk_j}^{\bfk_a \bfk_b}= \bra ij || ab \ket \frac{c_{\bfk_i \bfk_j}^{\bfk_a \bfk_b} }{c_{\bf 0} },
\end{equation}
Recalling that this set of $k$-points is bounded by a maximum kinetic energy, and thus has a basis set size ($M$) associated with it as defined at the start of \refsec{sec_FCIQMC}. The upper limit on the virtual space sum is therefore modified,
\begin{equation}
E_{\text{corr},k} \left(M \right)=\sum_{ij}^{\text{occ}} \sum_{ab}^{M} \chi_{\bfk_i \bfk_j}^{\bfk_a \bfk_b}.
\label{ecorrk}
\end{equation}
We propose to regroup these energy contributions according to their arrangement in reciprocal space and use the behaviour of the coefficients to provide an estimate of the complete basis set limit. In doing so, we will construct energies of an effective basis sets of a smaller size, which can be viewed as groups of plane waves that lie on concentric spheres in reciprocal space, from a single large basis set calculation. Via extrapolation from all of these smaller effective basis set sizes, an estimate for the CBS energy can then be obtained without the need for further calculations.

In order to regroup these coefficients, we construct \emph{masking functions} $P$, equal to either 0 or 1, to effectively remove some of the terms in the sum in \refeq{ecorrk},
\begin{equation}
E_{\text{corr},\text{eff}} \left(M \right)=\sum_{ij}^{\text{occ}} \sum_{ab}^{M} \chi_{\bfk_i \bfk_j}^{\bfk_a \bfk_b}  P \left(\bfk_a,\bfk_b\right).
\end{equation}
out of spherical step functions,
\begin{equation}
\Theta \left(k-k_c\right) = \left\{
\begin{array}{ll}
1, & | \bfk | \leq k_c \\
0, & \text{otherwise}
\end{array}
\right. 
\end{equation}
such that these step functions, and hence the masking functions, have associated with them a kinetic energy cutoff analogous to the original energy cutoff for the calculation. These are then multiplied by the coefficients of the projected energy and re-summed at different new energy cutoffs that are smaller than the original kinetic energy cutoff that the simulation was performed.

In contrast to the original basis sets defined here, we construct these effective basis sets by using cutoffs based on the momentum transfer of each excitation. For $ij\rightarrow ab$, this is given by,
\begin{equation}
\bfk_a=\bfk_i+\bfg \quad;\quad \bfk_b=\bfk_j-\bfg,
\end{equation}
or, equivalently,
\begin{equation}
\bfk_a=\bfk_j-\bfg^\prime \quad;\quad \bfk_b=\bfk_i+\bfg^\prime,
\end{equation}
due to conservation of momentum ($\bfk_i+\bfk_j=\bfk_a+\bfk_b$). The masking function we have found to be most successful is,
\begin{equation}
\begin{split}
P_g\left(\bfg, \bfg^\prime ; M_g  \right) &=\Theta \left(g-g_c\right) +\Theta \left(g^\prime-g_c\right) \\
&-\Theta \left(g-g_c\right)\Theta \left(g^\prime-g_c\right),
\end{split}
\end{equation}
which denotes the union of the set of k-points enclosed by two spheres of radius $g_c$ centred on the arguments of the function, $\bfg$ and $\bfg^\prime$. This leads to an expression for the energies of an \emph{effective basis set} size $M^\prime$, due to different \emph{effective cutoffs} $g_c$,
\begin{equation}
\begin{split}
E_{\text{corr},\text{eff}} \left(M, M^\prime \right)&=\sum_{ij}^{\text{occ}} \sum_{ab}^{M} \chi_{\bfk_i \bfk_j}^{\bfk_a \bfk_b} \\
&\times P_g\left(\bfk_i-\bfk_a,\bfk_j-\bfk_a ; M^\prime \right),
\end{split}
\end{equation}
This \emph{effective basis set} size $M^\prime$, denotes a truncated basis which encompasses twice the number of $k$-points enclosed by a sphere of radius $g_c$ centred at the $\Gamma$-point. The behaviour of the energies due to these effective basis sets, in the limit that $M$ is large enough to completely enclose all possible excitations of length $g_c$, is also of form $1/M^\prime$. However, convergence on this linear behaviour is much faster and we are able to compute approximate CBS limit estimates from only one calculation. We therefore call this \emph{single point extrapolation}.

In \reffig{proje_graph} we show an example of this technique being used to compute a CBS limit estimate from $M=1850$. This agrees well with the CBS estimate from a normal extrapolation as in \refsec{BSSE}, however this single point extrapolation scheme is found to converge at lower basis set sizes $M$ (\reffig{proje_graph}) as well as providing a reliable extrapolation at each point. Errors in this technique arise from coefficient relaxation, due to the changing value of $\chi_{\bfk_i \bfk_j}^{\bfk_a \bfk_b}$ as $M$ is varied. However, we find that there is cancellation of errors between the approximate effective basis energies and the resulting $1/M^\prime$ gradient, and so convergence is rapid.

\begin{table*}

\begin{tabular*}{0.85\textwidth}{@{\extracolsep{\fill}}  c  c  c  c  c c c c c }
 $r_s$ (a.u.) & $N$ & $E_\text{corr}$ (a.u.) & $N_w$ & $M$ & $S$ (a.u.) & $n_\text{add}$ & $E_\text{corr,DMC}$ (a.u.) & CPU time (corehours)\\
  \hline
  \hline
 \multirow{3}{*}{0.5} & 14 & -0.5959(7) & $10^6$ & 1850 & 0.0 & 3 &  - & 200 \\
& 38 & -1.849(1) & $10^8$ & 922 & 0.1 &3&  - & 4,000\\
& 54 & -2.435(7) & $10^7$ & 922 & 0.1 &3& -2.387(2) &4,000 \\
\hline
 \multirow{3}{*}{1.0} & 14 & -0.5316(4) & $10^7$ & 1850 & 0.0 &3&  - & 2,500\\
& 38 & -1.590(1) & $10^8$ & 922 & 0.1 &3&  - & 8,000 \\
& 54 & -2.124(3) & $10^{8}$ & 922 & 0.1 &3&  -2.125(2) & 6,000  \\
\hline
 \multirow{3}{*}{2.0} & 14 & -0.444(1) & $10^7$ & 1850 & -0.2 &3&  - & 2,500\\
& 38 & -1.225(2) & $10^9$ & 922 & 0.1 &3&  - & 16,000 \\
\hline
 \multirow{1}{*}{5.0} & 14 & -0.307(1) & $10^{10}$ & 778 & -0.2 &3&  - & 40,000 \\
  \hline
\end{tabular*}
  \caption{\iFCIQMC complete basis set total correlation energies for a variety of $N$ and $r_s$, estimated using the projected energy \emph{single point extrapolation} technique described in the text. The source of the error estimate is stochastic error. The results compare well with DMC results obtained by R\'{\i}os \emph{et al.} \cite{Rios}, which are comparable to those found by Kwon \emph{et al.}\cite{Kwon1998}. For further discussion of this comparison, see Ref. \onlinecite{UEGPaper1}.}
  
  \label{table2}
\end{table*}

It is worthwhile pointing out that this scheme is a marked contrast to the extrapolation scheme mentioned before in \refsec{BSSE} where separate calculations, each variationally the lowest energy achievable in a one-particle basis set, were used to extrapolate to the CBS limit. It is more common in the quantum chemical literature to take this previous approach and as such \emph{single point extrapolation} goes against the prevailing literature. We accept that these results will only be entirely accurate in the CBS limit of $M\rightarrow\infty$, but note that this limit can be found systematically, and as such our results should be treated as CBS estimates. Furthermore, in the HEG, there is no orbital relaxation in the Hartree-Fock orbitals as $M$ increases, since they are exact. We nonetheless believe this approach to be a reasonable one to take in plane wave systems, where there is great flexibility in the basis set size, slow basis set convergence, and in particular, is the most practical approach in terms of computational cost for the systems studied. Moreover, in real solid state systems, it is currently necessary to use an auxiliary basis set to find results at the complete basis set limit\cite{Kresse2009}.

In FCIQMC, the benefits of using a single point extrapolation are substantial, due to the effects of the initiator approximation. The approximate form of the curve in \reffig{proje_graph} appears to converge very rapidly with respect to walker number, in particular allowing for an estimation of where the linear regime begins at very low computational cost. Furthermore, once a basis size has been chosen to perform a single point extrapolation from, the initiator error in the coefficients should be consistent for all effective basis set sizes, somewhat mitigating errors due to change in initiator error across different basis sizes in the more traditional extrapolation scheme.

It is also possible to calculate an estimate for the CBS limit on the fly during a simulation, when the region of linear $1/M^\prime$ behaviour is known, allowing for the production of an easily computable, rapidly convergent, stochastic correction. Furthermore, it is possible to probe the convergence in a variety of orbital subspaces using different masking functions, which may be useful in the future to help understand the nature of the initiator approximation from the point of view of the one-particle basis set. Taken together, these extend the practical use of such a technique, however extensive study of this is beyond the scope of this paper.

We therefore conclude by presenting a set of correlation energies in \reftab{table2}. We are confident that all 14 electron results are free from both initiator and basis set incompleteness error, although we cannot rule out this possibility for the higher electron numbers. Convergence of the $E_g$-cutoff extrapolation at a range of $r_s$-values is shown in \reffig{rs_splitproje} and \reftab{table2}. $M$ is assumed to be large enough for the single-point extrapolation to remove the basis set incompleteness error. This is supported by agreement (within stochastic error bars) with energies presented in \refsec{BSSE}, in which the CBS limit is achieved from a conventional $1/M$ extrapolation, and with values published in a previous paper on the 54-electron problem. They also compare well with the most accurate DMC calculations to date\cite{Rios,Kwon1998}, in particular yielding a lower energy at $N=54$ at $r_s=0.5$, again consistent with our previous paper (Ref. \onlinecite{UEGPaper1}). We also note that computational cost of these new CBS \iFCIQMC energies is approximately 100 times smaller than previously\footnote{This is predominantly from the new form of extrapolation used. We note however that the amount of time taken to converge the 54-electron, $r_s=1.0$, $M=922$ energy by the fixed shift strategy was of the same order as the amount of time taken statistics accumulation for the same basis set size in Ref. \onlinecite{UEGPaper1}.}.

\section{Concluding remarks}

In this paper we have applied \iFCIQMC, to the simulation cell HEG at a variety of system sizes, $N=14,38$ and $54$ electrons, over a range of correlation strengths $0.5 \leq r_s \leq 5.0$~a.u.

We develop the use of a \emph{fixed shift strategy} to examine convergence of the calculations to the large walker limit. The \iFCIQMC method has associated with it two sources of error when trying to calculate FCI accuracy energies. These are stochastic error, arising from the evolution of the discretized wavefunction coefficients through imaginary time, and initiator error, arising from using a finite number of walkers for the simulation. We investigated these two sources of error and showed that, with a small modification to the current algorithm, they can be independently reduced, removed or quantified systematically. In so doing we also gave an explanation of the internal parameters within an \iFCIQMC simulation and how these can be optimized for computational efficiency. 

Making use of the easily-tunable basis set of the HEG, we demonstrated that the basis set scaling for the very large basis sets of a weakly correlated system ($N=14$, $r_s=1.0$~a.u.) was approximately $\mathcal{O} \left[ M \right]$. We could find no evidence in the very high basis set limit for an exponential scaling, as previous studies of molecular systems have identified\cite{EApaper,c2paper}, although note that our conclusion would almost certainly change with system size and strength of correlation.

Finally, we applied the newly-developed \emph{single point extrapolation} for the projected energy, which uses information from a single large-basis-set calculation to extrapolate to the complete basis set limit, and successfully yield complete basis set energies for a range of $N$ and $r_s$. We note that in combination with the fixed shift strategy, this leads to a 100 fold saving in computational cost in producing complete basis set energies for the 54 electron problem\cite{UEGPaper1}.

In so doing we hope that we have demonstrated both that the HEG is a versatile and useful model system, providing benchmarks for the future application of quantum chemical techniques, and also more rigorously analyzed some of the open methodological questions surrounding \iFCIQMCbracket.

\acknowledgments

This work was supported by a grant from the Distributed European Infrastructure for Supercomputing Applications under their Extreme Computing Initiative. For funding, the authors acknowledge Trinity College (GHB) and EPSRC (JJS, AA), and we are grateful to A Gr\"uneis, J S Spencer and A J W Thom for discussions.


\end{document}